\documentclass[onecolumn]{article}%
\usepackage{amsmath}
\usepackage{graphicx}
\usepackage{amsfonts}
\usepackage{amssymb}
\usepackage{fullpage}%
\newtheorem{theorem}{Theorem}

\newtheorem{algorithm}[theorem]{Algorithm}

\newtheorem{corollary}[theorem]{Corollary}

\newtheorem{definition}[theorem]{Definition}

\newtheorem{lemma}[theorem]{Lemma}

\newtheorem{proposition}[theorem]{Proposition}

\newenvironment{proof}[1][Proof]{\textbf{#1.} }{\ \rule{0.5em}{0.5em}}
\begin{document}

\title{Quantum Search of Spatial Regions}
\author{Scott Aaronson\thanks{Email: aaronson@ias.edu. \ This work was mostly done
while the author was a PhD student at UC Berkeley, supported by an NSF
Graduate Fellowship.}\\Institute for Advanced Study, Princeton
\and Andris Ambainis\thanks{Email: ambainis@iqc.ca. \ Supported by an IQC
University Professorship and by CIAR. \ This work was mostly done while the
author was at the University of Latvia.}\\University of Waterloo}
\date{}
\maketitle

\begin{abstract}
Can Grover's algorithm speed up search of a physical region---for example a
$2$-D grid of size $\sqrt{n}\times\sqrt{n}$? \ The problem is that $\sqrt{n}%
$\ time seems to be needed for each query, just to move amplitude across the
grid. \ Here we show that this problem can be surmounted, refuting a claim to
the contrary by Benioff. \ In particular, we show how to search a
$d$-dimensional hypercube in time $O(\sqrt{n})$ for $d\geq3$, or $O(\sqrt
{n}\log^{5/2}n)$\ for $d=2$.\ \ More generally, we introduce a model of
\textit{quantum query complexity on graphs}, motivated by fundamental physical
limits on information storage, particularly the holographic principle\ from
black hole thermodynamics. \ Our results in this model include almost-tight
upper and lower bounds for many search tasks; a generalized algorithm that
works for any graph with good expansion properties, not just hypercubes; and
relationships among several notions of `locality' for unitary matrices acting
on graphs. \ As an application of our results, we give an $O(\sqrt{n}%
)$-qubit\ communication protocol for the disjointness problem, which improves
an\ upper bound of H\o yer and de Wolf\ and matches a lower bound of Razborov.

\end{abstract}

\section{Introduction\label{INTRO}}

The goal of Grover's quantum search algorithm \cite{grover,grover:prl}\ is to
search an `unsorted database' of size $n$ in a number of queries proportional
to $\sqrt{n}$. \ Classically, of course, order $n$ queries are needed. \ It is
sometimes asserted that, although the speedup of Grover's algorithm\ is only
quadratic, this speedup is \textit{provable}, in contrast to the exponential
speedup of Shor's factoring algorithm \cite{shor}. \ But is that really true?
\ Grover's algorithm is typically imagined as speeding up combinatorial
search---and we do not know whether every problem in $\mathsf{NP}$ can be
classically solved quadratically faster than the \textquotedblleft
obvious\textquotedblright\ way, any more than we know whether factoring is in
$\mathsf{BPP}$.

\begin{figure}
[ptb]
\begin{center}
\includegraphics[
height=113.5pt,
width=243.0625pt
]%
{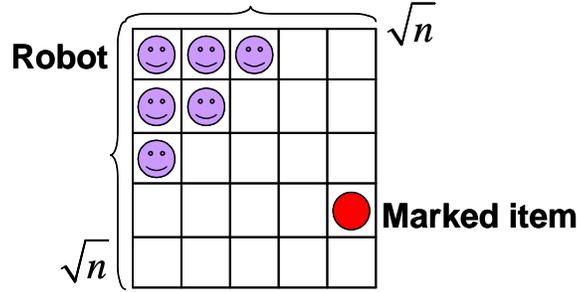}%
\caption{A quantum robot, in a superposition over locations, searching for a
marked item on a 2D grid of size $\sqrt{n}\times\sqrt{n}$.}%
\label{gridfig}%
\end{center}
\end{figure}

But could Grover's algorithm speed up search of a \textit{physical region}?
\ Here the basic problem, it seems to us, is the time needed for signals to
travel across the region. \ For if we are interested in the fundamental limits
imposed by physics, then we should acknowledge that the speed of light is
finite, and that a bounded region of space can store only a finite amount of
information, according to the holographic principle\ \cite{bousso}. \ We
discuss the latter constraint in detail in Section \ref{PHYS}; for now, we say
only that it suggests a model in which a `quantum robot' occupies a
superposition over finitely many locations, and moving the robot from one
location to an adjacent one takes unit time. \ In such a model, the time
needed to search a region could depend critically on its spatial layout. \ For
example,\ if the $n$ entries are arranged on a line, then even to move the
robot from one end to the other takes $n-1$ steps. \ But what if the entries
are arranged on, say, a $2$-dimensional square grid (Figure \ref{gridfig})?%

\subsection{Summary of Results\label{SUMMARY}}

This paper gives the first systematic treatment of quantum search of spatial
regions, with `regions' modeled as connected graphs. \ Our main result is
positive: we show that a quantum robot can search a $d$-dimensional hypercube
with $n$ vertices for a unique marked vertex in time $O\left(  \sqrt{n}%
\log^{3/2}n\right)  $\ when $d=2$, or $O\left(  \sqrt{n}\right)  $ when
$d\geq3$. \ This matches (or in the case of $2$ dimensions, nearly matches)
the $\Omega\left(  \sqrt{n}\right)  $\ lower bound for quantum search, and
supports the view that Grover search of a physical region presents no problem
of principle.\ Our basic technique is divide-and-conquer; indeed, once the
idea is pointed out, an upper bound of $O\left(  n^{1/2+\varepsilon}\right)
$\ follows readily. \ However, to obtain the tighter bounds is more difficult;
for that we use the amplitude-amplification framework of Grover
\cite{grover:framework} and Brassard et al. \cite{bhmt}.

Section \ref{GRID} presents the main results;\ Section \ref{MULTIPLE}\ shows
further that, when there are $k$ or more marked vertices, the search time
becomes $O\left(  \sqrt{n}\log^{5/2}n\right)  $\ when $d=2$, or $\Theta\left(
\sqrt{n}/k^{1/2-1/d}\right)  $\ when $d\geq3$.\ \ Also, Section \ref{IRREG}
generalizes our algorithm to arbitrary graphs that have `hypercube-like'
expansion properties. \ Here the best bounds we can achieve are $\sqrt
{n}2^{O\left(  \sqrt{\log n}\right)  }$\ when $d=2$, or $O\left(  \sqrt
{n}\operatorname*{polylog}n\right)  $ when $d>2$\ (note that $d$ need not be
an integer). \ Table \ref{sumresults}\ summarizes the results.

\begin{table}[ptb]
\label{sumresults}
\begin{tabular}
[c]{c|cc}
& $d=2$ & $d>2$\\\hline
\multicolumn{1}{r|}{Hypercube, $1$ marked item} &
\multicolumn{1}{|l}{$O\left(  \sqrt{n}\log^{3/2}n\right)  $} &
\multicolumn{1}{l}{$\Theta\left(  \sqrt{n}\right)  $}\\
\multicolumn{1}{r|}{Hypercube, $k$ or more marked items} &
\multicolumn{1}{|l}{$O\left(  \sqrt{n}\log^{5/2}n\right)  $} &
\multicolumn{1}{l}{$\Theta\left(  \frac{\sqrt{n}}{k^{1/2-1/d}}\right)  $}\\
\multicolumn{1}{r|}{Arbitrary graph, $k$ or more marked items} &
\multicolumn{1}{|l}{$\sqrt{n}2^{O\left(  \sqrt{\log n}\right)  }$} &
\multicolumn{1}{l}{$\widetilde{\Theta}\left(  \frac{\sqrt{n}}{k^{1/2-1/d}%
}\right)  $}%
\end{tabular}
\caption{Upper and lower bounds for quantum search on a $d$-dimensional graph
given in this paper. \ The symbol $\widetilde{\Theta}$\ means that the upper
bound includes a polylogarithmic term. \ Note that, if $d=2$, then
$\Omega\left(  \sqrt{n}\right)  $\ is always a lower bound, for any number of
marked items.}%
\end{table}

Section \ref{APPL}\ shows, as an unexpected application of our search
algorithm, that the quantum communication complexity of the well-known
\textit{disjointness problem} is $O\left(  \sqrt{n}\right)  $. \ This improves
an $O\left(  \sqrt{n}c^{\log^{\ast}n}\right)  $\ upper bound of H\o yer and de
Wolf \cite{hoyerdewolf},\ and matches the $\Omega\left(  \sqrt{n}\right)
$\ lower bound of Razborov \cite{razborov:cc}.

The rest of the paper is about the formal model that underlies our results.
\ Section \ref{PHYS} sets the stage for this model, by exploring the ultimate
limits on information storage imposed by properties of space and time. \ This
discussion\ serves only to motivate our results; thus, it can be safely
skipped by readers unconcerned with the physical universe. \ In Section
\ref{MODEL}\ we define \textit{quantum query algorithms on graphs}, a model
similar to quantum query algorithms as defined by Beals et al. \cite{bbcmw},
but with the added requirement that unitary operations be `local' with respect
to some graph. \ In Section \ref{LOCAL}\ we address the difficult question,
which also arises in work on quantum random walks \cite{aakv}\ and quantum
cellular automata \cite{watrous:ca}, of what `local' means. Section
\ref{GENERAL}\ proves general facts about our model, including an upper bound
of $O\left(  \sqrt{n\delta}\right)  $\ for the time needed to search any graph
with diameter $\delta$, and a proof (using the hybrid argument of Bennett et
al. \cite{bbbv}) that this upper bound is tight for certain graphs. \ We
conclude in Section \ref{OPEN}\ with some open problems.

\subsection{Related Work\label{PREV}}

In a paper on `Space searches with a quantum robot,' Benioff
\cite{benioff:robot}\ asked whether Grover's algorithm\ can speed up search of
a physical region, as opposed to a combinatorial search space. \ His answer
was discouraging: for a $2$-D grid of size $\sqrt{n}\times\sqrt{n}$, Grover's
algorithm is no faster than classical search. \ The reason is that, during
each of the $\Theta\left(  \sqrt{n}\right)  $ Grover iterations, the algorithm
must use order $\sqrt{n}$\ steps just to travel across the grid and return to
its starting point for the diffusion step. \ On the other hand, Benioff noted,
Grover's algorithm does yield some speedup for grids of dimension $3$ or
higher, since those grids have diameter less than $\sqrt{n}$.

Our results show that Benioff's claim is mistaken: by using Grover's algorithm
more carefully, one can search a $2$-D grid for a single marked vertex in
$O\left(  \sqrt{n}\log^{3/2}n\right)  $ time. \ To us this illustrates why one
should not assume an algorithm is optimal on heuristic grounds. \ Painful
experience---for example, the \textquotedblleft obviously
optimal\textquotedblright\ $O\left(  n^{3}\right)  $\ matrix multiplication
algorithm \cite{strassen}---is what taught computer scientists to see the
proving of lower bounds as more than a formality.

Our setting is related to that of quantum random walks on graphs
\cite{aakv,ccdfgs,cfg,skw}. \ In an earlier version of this paper, we asked
whether quantum walks might yield an alternative spatial search algorithm,
possibly even one that outperforms our divide-and-conquer algorithm.
\ Motivated by this question,\ Childs and Goldstone \cite{cg} managed to show
that in the continuous-time setting, a quantum walk can search a
$d$-dimensional hypercube for a single marked vertex in time $O\left(
\sqrt{n}\log n\right)  $ when $d=4$, or $O\left(  \sqrt{n}\right)  $ when
$d\geq5$. \ Our algorithm was still faster in $3$ or fewer dimensions\ (see
Table \ref{bounds}). \ Subsequently, however, Ambainis, Kempe, and Rivosh
\cite{akr}\ gave an algorithm based on a discrete-time quantum walk, which was
as fast as ours in $3$ or more dimensions, and faster in $2$ dimensions. \ In
particular, when $d=2$\ their algorithm used only $O\left(  \sqrt{n}\log
n\right)  $\ time to find a unique marked vertex. \ Childs and Goldstone
\cite{cg2} then gave a continuous-time quantum walk algorithm with the same
performance, and related this algorithm to properties of the Dirac equation.
\ It is still open whether $O\left(  \sqrt{n}\right)  $\ time is achievable in
$2$ dimensions.

\begin{table}[ptb]
\label{bounds}
\begin{tabular}
[c]{l|cccc}
& $d=2$ & $d=3$ & $d=4$ & $d\geq5$\\\hline
This paper & \multicolumn{1}{|l}{$O\left(  \sqrt{n}\log^{3/2}n\right)  $} &
\multicolumn{1}{l}{$O\left(  \sqrt{n}\right)  $} &
\multicolumn{1}{l}{$O\left(  \sqrt{n}\right)  $} &
\multicolumn{1}{l}{$O\left(  \sqrt{n}\right)  $}\\
\cite{cg} & \multicolumn{1}{|l}{$O\left(  n\right)  $} &
\multicolumn{1}{l}{$O\left(  n^{5/6}\right)  $} & \multicolumn{1}{l}{$O\left(
\sqrt{n}\log n\right)  $} & \multicolumn{1}{l}{$O\left(  \sqrt{n}\right)  $}\\
\cite{akr,cg2} & \multicolumn{1}{|l}{$O\left(  \sqrt{n}\log n\right)  $} &
\multicolumn{1}{l}{$O\left(  \sqrt{n}\right)  $} &
\multicolumn{1}{l}{$O\left(  \sqrt{n}\right)  $} &
\multicolumn{1}{l}{$O\left(  \sqrt{n}\right)  $}%
\end{tabular}
\caption{Time needed to find a unique marked item in a $d$-dimensional
hypercube, using the divide-and-conquer algorithms of this paper, the original
quantum walk algorithm of Childs and Goldstone \cite{cg}, and the improved
walk algorithms of Ambainis, Kempe, and Rivosh \cite{akr}\ and Childs and
Goldstone \cite{cg2}.}%
\end{table}

Currently, the main drawback of the quantum walk approach is that all analyses
have relied heavily on symmetries in the underlying graph. \ If even minor
`defects' are introduced, it is no longer known how to upper-bound the running
time. \ By contrast, the analysis of our divide-and-conquer algorithm is
elementary, and does not depend on eigenvalue bounds. \ We can therefore show
that the algorithm works for any graphs with sufficiently good expansion properties.

Childs and Goldstone \cite{cg}\ argued that the quantum walk approach has the
advantage of requiring fewer auxiliary qubits than the divide-and-conquer
approach. \ However, the need for many qubits\ was an artifact of how we
implemented the algorithm in a previous version of the paper. \ The current
version uses only \textit{one} qubit.

\section{The Physics of Databases\label{PHYS}}

Theoretical computer science generally deals with the limit as some resource
(such as time or memory) increases to infinity. \ What is not always
appreciated is that, as the resource bound increases, physical constraints may
come into play that were negligible at `sub-asymptotic'\ scales.\ \ We believe
theoretical computer scientists ought to know something about such
constraints, and to account for them when possible. \ For if the constraints
are ignored on the ground that they \textquotedblleft never matter in
practice,\textquotedblright\ then the obvious question arises: why use
asymptotic analysis in the first place, rather than restricting attention to
those instance sizes that occur in practice?

A constraint of particular interest for us is the \textit{holographic
principle} \cite{bousso},\ which arose from black-hole thermodynamics. \ The
principle states that the information content of any spatial region is
upper-bounded by its \textit{surface area} (not volume), at a rate of one bit
per Planck area, or about $1.4\times10^{69}$\ bits per square meter.
\ Intuitively, if one tried to build a spherical hard disk with mass density
$\upsilon$, one could not keep expanding it forever. \ For as soon as the
radius reached the Schwarzschild bound of $r=\sqrt{3/\left(  8\pi
\upsilon\right)  }$ (in Planck units, $c=G=\hbar=k=1$), the hard disk would
collapse to form a black hole, and thus its contents would be irretrievable.

Actually the situation is worse than that: even a \textit{planar} hard disk of
constant mass density would collapse to form a black hole once its radius
became sufficiently large, $r=\Theta\left(  1/\upsilon\right)  $. \ (We assume
here that the hard disk is disc-shaped. \ A linear or $1$-D hard disk could
expand indefinitely without collapse.) \ It is possible, though, that a hard
disk's information content could asymptotically exceed its mass. \ For
example, a black hole's mass is proportional to the radius of its event
horizon, but the entropy is proportional to the \textit{square} of the radius
(that is, to the surface area). \ Admittedly, inherent difficulties with
storage and retrieval make a black hole horizon less than ideal as a hard
disk. \ However, even a weakly-gravitating system could store information at a
rate asymptotically exceeding its mass-energy. \ For instance, Bousso
\cite{bousso}\ shows that an enclosed ball of radiation with radius $r$ can
store\ $n=\Theta\left(  r^{3/2}\right)  $ bits, even though its energy grows
only as $r$. \ Our results in Section\ \ref{SCATTERED}\ will imply that a
quantum robot could (in principle!) search such a `radiation disk' for a
marked item in time $O\left(  r^{5/4}\right)  =O\left(  n^{5/6}\right)  $.
\ This is some improvement over the trivial $O\left(  n\right)  $ upper bound
for a $1$-D hard disk, though it falls short of the desired $O\left(  \sqrt
{n}\right)  $.

In general, if $n=r^{c}$ bits are scattered throughout a $3$-D ball of radius
$r$ (where $c\leq3$\ and the bits' locations are known), we will show in
Theorem \ref{scatterthm}\ that the time needed to search for a `$1$'\ bit
grows as $n^{1/c+1/6}=r^{1+c/6}$ (omitting logarithmic factors). \ In
particular, if $n=\Theta\left(  r^{2}\right)  $\ (saturating the holographic
bound), then the time grows as $n^{2/3}$ or $r^{4/3}$. \ To achieve a search
time of $O\left(  \sqrt{n}\operatorname*{polylog}n\right)  $, the bits would
need to be concentrated on a $2$-D surface.

Because of the holographic principle, we see that it is not only quantum
mechanics that yields a $\Omega\left(  \sqrt{n}\right)  $\ lower bound on the
number of steps needed for unordered search. \ If the items to be searched are
laid out spatially, then general relativity in $3+1$\ dimensions independently
yields the same bound, $\Omega\left(  \sqrt{n}\right)  $, up to a constant
factor.\footnote{Admittedly, the holographic principle is part of quantum
gravity and not general relativity \textit{per se}. \ All that matters for us,
though, is that the principle seems logically independent of
quantum-mechanical linearity, which is what produces the \textquotedblleft
other\textquotedblright\ $\Omega\left(  \sqrt{n}\right)  $\ bound.}
\ Interestingly, in $d+1$\ dimensions the relativity bound would be
$\Omega\left(  n^{1/\left(  d-1\right)  }\right)  $, which\ for $d>3$\ is
weaker than the quantum mechanics bound. \ Given that our two fundamental
theories yield the same lower bound, it is natural to ask whether that bound
is tight. \ The answer seems to be that it is \textit{not} tight, since (i)
the entropy on a black hole horizon is not efficiently accessible\footnote{In
the case of a black hole horizon, waiting for the bits to be emitted as
Hawking radiation---as recent evidence suggests that they are \cite{sv}%
---takes time proportional to $r^{3}$,\ which is much too long.}, and (ii)
weakly-gravitating systems are subject to the \textit{Bekenstein bound}
\cite{bekenstein}, an even stronger entropy constraint than the holographic bound.

Yet it is still of basic interest to know whether $n$ bits in a radius-$r$
ball can be searched in time $o\left(  \min\left\{  n,r\sqrt{n}\right\}
\right)  $---that is, whether it is possible to do \textit{anything} better
than either brute-force quantum search (with the drawback pointed out by
Benioff \cite{benioff:robot}), or classical search. \ Our results show that it
is possible.

From a physical point of view, several questions naturally arise: (1) whether
our complexity measure is realistic; (2) how to account for time dilation; and
(3) whether given the number of bits we are imagining, cosmological bounds are
also relevant. \ Let us address these questions in turn.

\begin{enumerate}
\item[(1)] One could argue that to maintain a `quantum database' of size $n$
requires $n$ computing elements (\cite{zalka}, though see also \cite{rg}).
\ So why not just exploit those elements to search the database in
\textit{parallel}? \ Then it becomes trivial to show that the search time is
limited only by the radius of the database, so the algorithms of this paper
are unnecessary. \ Our response is that, while there might be $n$ `passive'
computing elements (capable of storing data), there might be many fewer
`active'\ elements, which we consequently wish to place in a superposition
over locations. \ This assumption seems physically unobjectionable. \ For a
particle (and indeed any object) really does have an indeterminate location,
not merely an indeterminate internal state (such as spin) \textit{at} some
location. \ We leave as an open problem, however, whether our assumption is
valid for specific quantum computer architectures such as ion traps.

\item[(2)] So long as we invoke general relativity, should we not also
consider the effects of time dilation? \ Those effects are indeed pronounced
near a black hole horizon. \ Again, though, for our upper bounds we will have
in mind systems\ far from the Schwarzschild limit, for which any time dilation
is by at most a constant factor independent of $n$.

\item[(3)] How do cosmological considerations affect our analysis? \ Bousso
\cite{bousso:vac}\ argues that, in a spacetime with positive cosmological
constant $\Lambda>0$, the total number of bits accessible to any one
experiment is at most $3\pi/\left(  \Lambda\ln2\right)  $, or roughly
$10^{122}$ given current experimental bounds \cite{perlmutter}\ on $\Lambda
$.\footnote{Also, Lloyd \cite{lloyd}\ argues that the total number of bits
accessible \textit{up till now} is at most the square of the number of Planck
times elapsed so far, or about $\left(  10^{61}\right)  ^{2}=10^{122}%
$.\ \ Lloyd's bound, unlike Bousso's, does not depend on $\Lambda$ being
positive. The numerical coincidence between the two bounds reflects the
experimental finding \cite{perlmutter,ryden}\ that we live in a transitional
era, when both $\Lambda$\ and \textquotedblleft dust\textquotedblright%
\ contribute significantly to the universe's net energy balance ($\Omega
_{\Lambda}\approx0.7$, $\Omega_{\operatorname*{dust}}\approx0.3$). \ In
earlier times dust (and before that radiation)\ dominated, and Lloyd's bound
was tighter. \ In later times $\Lambda$\ will dominate, and Bousso's bound
will be tighter. \ \textit{Why} we should live in such a transitional era is
unknown.} \ Intuitively, even if the universe is spatially infinite, most of
it recedes too quickly from any one observer to be harnessed as computer memory.

One response to this result is to assume an idealization in which $\Lambda
$\ vanishes, although Planck's constant $\hbar$ does not vanish. \ As
justification, one could argue that without the idealization $\Lambda=0$,
\textit{all} asymptotic bounds in computer science are basically fictions.
\ But perhaps a better response is to accept the $3\pi/\left(  \Lambda
\ln2\right)  $ bound, and then ask how close one can come to
\textit{saturating} it in different scenarios. \ Classically, the maximum
number of bits that can be searched is, in a crude
model\footnote{Specifically, neglecting gravity and other forces that could
counteract the effect of $\Lambda$.}, actually proportional to $1/\sqrt
{\Lambda}\approx10^{61}$ rather than $1/\Lambda$. \ The reason is that if a
region had much more than $1/\sqrt{\Lambda}$\ bits, then after $1/\sqrt
{\Lambda}$\ Planck times---that is, about $10^{10}$\ years, or roughly the
current age of the universe---most of the region would have receded beyond
one's cosmological horizon. \ What our results suggest is that, using a
quantum robot, one could come closer to saturating the cosmological
bound---since, for example, a $2$-D region of size $1/\Lambda$ can be searched
in time $O\left(  \frac{1}{\sqrt{\Lambda}}\operatorname*{polylog}\frac
{1}{\sqrt{\Lambda}}\right)  $. \ How anyone could \textit{prepare} a database
of size much greater than $1/\sqrt{\Lambda}$\ remains unclear, but if such a
database existed, it could be searched!
\end{enumerate}

\section{The Model\label{MODEL}}

Much of what is known about the power of quantum computing comes from the
\textit{black-box} or \textit{query} model
\cite{ambainis,bbcmw,bbbv,grover,shor}, in which one counts only the number of
queries to an oracle, not the number of computational steps. \ We will take
this model as the starting point for a formal definition of quantum robots.
\ Doing so will focus attention on our main concern: how much harder is it to
evaluate a function when its inputs are spatially separated? \ As it turns
out, all of our algorithms \textit{will} be efficient as measured by the
number of gates and auxiliary qubits needed to implement them.

For simplicity, we assume that a robot's goal is to evaluate a Boolean
function $f:\left\{  0,1\right\}  ^{n}\rightarrow\left\{  0,1\right\}  $,
which could be partial or total. \ A `region of space' is a connected
undirected graph $G=\left(  V,E\right)  $\ with vertices $V=\left\{
v_{1},\ldots,v_{n}\right\}  $. \ Let $X=x_{1}\ldots x_{n}\in\left\{
0,1\right\}  ^{n}$\ be an input to $f$; then each bit\ $x_{i}$\ is available
only at vertex $v_{i}$. \ We assume the robot knows $G$ and the vertex labels
in advance, and so is ignorant only of the $x_{i}$ bits. \ We thus sidestep a
major difficulty for quantum walks \cite{aakv}, which is how to ensure that a
process on an unknown graph is unitary.

At any time, the robot's state has the form%
\[
\sum\alpha_{i,z}\left|  v_{i},z\right\rangle \text{.}%
\]
Here $v_{i}\in V$\ is a vertex, representing the robot's location; and $z$ is
a bit string (which can be arbitrarily long), representing the robot's
internal configuration. \ The state evolves via an alternating sequence of
$T$\ algorithm\ steps and $T$ oracle steps:%
\[
U^{\left(  1\right)  }\rightarrow O^{\left(  1\right)  }\rightarrow U^{\left(
1\right)  }\rightarrow\cdots\rightarrow U^{\left(  T\right)  }\rightarrow
O^{\left(  T\right)  }\text{.}%
\]
An oracle step $O^{\left(  t\right)  }$ maps each basis state $\left|
v_{i},z\right\rangle $\ to $\left|  v_{i},z\oplus x_{i}\right\rangle $, where
$x_{i}$\ is exclusive-OR'ed into the first bit of $z$. \ An algorithm step
$U^{\left(  t\right)  }$ can be any unitary matrix that (1) does not depend on
$X$, and (2) acts `locally' on $G$. \ How to make the second condition precise
is the subject of Section \ref{LOCAL}.

The initial state of the algorithm is $\left\vert v_{1},0\right\rangle $.
\ Let $\alpha_{i,z}^{\left(  t\right)  }\left(  X\right)  $\ be the amplitude
of $\left\vert v_{i},z\right\rangle $\ immediately after the $t^{th}$\ oracle
step; then the algorithm succeeds with probability $1-\varepsilon$\ if%
\[
\sum_{\left\vert v_{i},z\right\rangle \,:\,z_{OUT}=f\left(  X\right)
}\left\vert \alpha_{i,z}^{\left(  T\right)  }\left(  X\right)  \right\vert
^{2}\geq1-\varepsilon
\]
for all inputs $X$, where $z_{OUT}$\ is a bit of $z$ representing the output.

\subsection{Locality Criteria\label{LOCAL}}

Classically, it is easy to decide whether a stochastic matrix acts
\textit{locally} with respect to a graph $G$:\ it does if it moves probability
only along the edges of $G$. \ In the quantum case, however, interference
makes the question much more subtle. \ In this section we propose three
criteria for whether a unitary matrix $U$ is local. \ Our algorithms will then
be implemented using the most restrictive of these criteria.

The first criterion we call \textit{Z-locality} (for zero): $U$ is Z-local if,
given any pair of non-neighboring vertices $v_{1},v_{2}$ in $G$, $U$ ``sends
no amplitude''\ from $v_{1}$\ to $v_{2}$; that is, the corresponding entries
in $U$ are all $0$. \ The second criterion, \textit{C-locality} (for
composability), says that this is not enough: not only must $U$ send amplitude
only between neighboring vertices, but it must be composed of a product of
commuting unitaries, each of which acts on a single edge. \ The third
criterion is perhaps the most natural one to a physicist: $U$ is
\textit{H-local} (for Hamiltonian) if it can be obtained by applying a
locally-acting, low-energy Hamiltonian for some fixed amount of time.\ \ More
formally, let $U_{i,z\rightarrow i^{\ast},z^{\ast}}$\ be the entry in the
$\left|  v_{i},z\right\rangle $\ column and $\left|  v_{i^{\ast}},z^{\ast
}\right\rangle $\ row of $U$.

\begin{definition}
$U$ is Z-local if $U_{i,z\rightarrow i^{\ast},z^{\ast}}=0$\ whenever $i\neq
i^{\ast}$\ and $\left(  v_{i},v_{i^{\ast}}\right)  $\ is not an edge of $G$.
\end{definition}

\begin{definition}
$U$ is C-local if the basis states can be partitioned into subsets
$P_{1},\ldots,P_{q}$\ such that

\begin{enumerate}
\item[(i)] $U_{i,z\rightarrow i^{\ast},z^{\ast}}=0$\ whenever $\left\vert
v_{i},z\right\rangle $\ and $\left\vert v_{i^{\ast}},z^{\ast}\right\rangle
$\ belong to distinct $P_{j}$'s, and

\item[(ii)] for each $j$, all basis states in $P_{j}$\ are either from the
same vertex or from two adjacent vertices.
\end{enumerate}
\end{definition}

\begin{definition}
$U$ is H-local if $U=e^{iH}$\ for some Hermitian $H$ with eigenvalues of
absolute value at most $\pi$, such that $H_{i,z\rightarrow i^{\ast},z^{\ast}%
}=0$\ whenever $i\neq i^{\ast}$\ and $\left(  v_{i},v_{i^{\ast}}\right)  $\ is
not an edge in $E$.
\end{definition}

If a unitary matrix is C-local, then it is also Z-local and H-local. \ For the
latter implication, note that any unitary $U$ can be written as $e^{iH}$\ for
some $H$ with eigenvalues of absolute value at most $\pi$. \ So we can write
the unitary $U_{j}$\ acting on each $P_{j}$ as $e^{iH_{j}}$; then since the
$U_{j}$'s\ commute,%
\[
\prod U_{j}=e^{i\sum H_{j}}\text{.}%
\]
Beyond that, though, how are the locality criteria related? \ Are they
approximately equivalent? \ If not, then does a problem's complexity in our
model ever depend on which criterion is chosen? \ Let us emphasize that these
questions are \textit{not} answered by, for example, the Solovay-Kitaev
theorem (see \cite{nc}), that an $n\times n$\ unitary matrix can be
approximated using a number of gates polynomial in $n$. \ For recall that the
definition of C-locality requires the edgewise operations to commute---indeed,
without that requirement, one could produce any unitary matrix at all. \ So
the relevant question, which we leave open, is whether any Z-local or H-local
unitary can be approximated by a product of, say, $O\left(  \log n\right)
$\ C-local unitaries. \ (A product of $O\left(  n\right)  $ such unitaries
trivially suffices, but that is far too many.)

\section{General Bounds\label{GENERAL}}

Given a Boolean function $f:\left\{  0,1\right\}  ^{n}\rightarrow\left\{
0,1\right\}  $, the quantum query complexity $Q\left(  f\right)  $,\ defined
by Beals et al. \cite{bbcmw}, is the minimum $T$ for which there exists a
$T$-query quantum algorithm that evaluates $f$ with probability at least $2/3$
on all inputs. \ (We will always be interested in the \textit{two-sided,
bounded-error} complexity, sometimes denoted $Q_{2}\left(  f\right)  $.)
\ Similarly, given a graph $G$ with $n$ vertices labeled $1,\ldots,n$, we let
$Q\left(  f,G\right)  $ be the minimum $T$ for which there exists a $T$-query
quantum robot on $G$ that evaluates $f$\ with probability $2/3$. \ Here we
require the algorithm steps to be C-local. \ One might also consider the
corresponding measures $Q^{Z}\left(  f,G\right)  $\ and $Q^{H}\left(
f,G\right)  $\ with Z-local and H-local steps respectively. \ Clearly
$Q\left(  f,G\right)  \geq Q^{Z}\left(  f,G\right)  $\ and $Q\left(
f,G\right)  \geq Q^{H}\left(  f,G\right)  $; we conjecture that all three
measures are asymptotically equivalent but were unable to prove this.

Let $\delta_{G}$\ be the diameter of $G$, and\ call $f$ \textit{nondegenerate}
if it depends on all $n$ input bits.

\begin{proposition}
\label{immed}For all $f,G$,

\begin{enumerate}
\item[(i)] $Q\left(  f,G\right)  \leq2n-3$.

\item[(ii)] $Q\left(  f,G\right)  \leq\left(  2\delta_{G}+1\right)  Q\left(
f\right)  $.

\item[(iii)] $Q\left(  f,G\right)  \geq Q\left(  f\right)  $.

\item[(iv)] $Q\left(  f,G\right)  \geq\delta_{G}/2$ if $f$ is nondegenerate.
\end{enumerate}
\end{proposition}

\begin{proof}
\quad

\begin{enumerate}
\item[(i)] Starting from the root, a spanning tree for $G$ can be traversed in
$2\left(  n-1\right)  -1$\ steps (there is no need to return to the root).

\item[(ii)] We can simulate a query in $2\delta_{G}$\ steps, by fanning out
from the start vertex $v_{1}$ and then returning. \ Applying a unitary at
$v_{1}$\ takes $1$ step.

\item[(iii)] Obvious.

\item[(iv)] There exists a vertex $v_{i}$ whose distance to $v_{1}$\ is at
least $\delta_{G}/2$, and $f$\ could depend on $x_{i}$.
\end{enumerate}
\end{proof}

We now show that the model is robust.

\begin{proposition}
\label{robust}For nondegenerate $f$, the following change $Q\left(
f,G\right)  $\ by at most a constant factor.

\begin{enumerate}
\item[(i)] Replacing the initial state $\left\vert v_{1},0\right\rangle $\ by
an arbitrary (known) $\left\vert \psi\right\rangle $.

\item[(ii)] Requiring the final state to be localized at some vertex $v_{i}%
$\ with probability at least $1-\varepsilon$, for a constant $\varepsilon>0$.

\item[(iii)] Allowing multiple algorithm steps between each oracle step (and
measuring the complexity by the number of algorithm steps).
\end{enumerate}
\end{proposition}

\begin{proof}

\begin{enumerate}
\item[(i)] We can transform $\left\vert v_{1},0\right\rangle $\ to $\left\vert
\psi\right\rangle $\ (and hence $\left\vert \psi\right\rangle $\ to
$\left\vert v_{1},0\right\rangle $) in $\delta_{G}=O\left(  Q\left(
f,G\right)  \right)  $ steps, by fanning out from $v_{1}$\ along the edges of
a minimum-height spanning tree.

\item[(ii)] Assume without loss of generality that $z_{OUT}$\ is accessed only
once, to write the output. \ Then after $z_{OUT}$\ is accessed, uncompute
(that is, run the algorithm backwards) to localize the final state at $v_{1}$.
\ The state can then be localized at any $v_{i}$\ in $\delta_{G}=O\left(
Q\left(  f,G\right)  \right)  $ steps. \ We can succeed with any constant
probability by repeating this procedure a constant number of times.

\item[(iii)] The oracle step $O$ is its own inverse, so we can implement a
sequence $U_{1},U_{2},\ldots$\ of algorithm steps as follows (where $I$ is the
identity):%
\[
U_{1}\rightarrow O\rightarrow I\rightarrow O\rightarrow U_{2}\rightarrow\cdots
\]

\end{enumerate}
\end{proof}

A function of particular interest is $f=\operatorname*{OR}\left(  x_{1}%
,\ldots,x_{n}\right)  $, which outputs $1$ if and only if $x_{i}=1$\ for some
$i$. \ We first give a general upper bound on $Q\left(  \operatorname*{OR}%
,G\right)  $\ in terms of the diameter of $G$. \ (Throughout the paper, we
sometimes omit floor and ceiling signs if they clearly have no effect on the asymptotics.)

\begin{proposition}
\label{upperdg}%
\[
Q\left(  \operatorname*{OR},G\right)  =O\left(  \sqrt{n\delta_{G}}\right)  .
\]

\end{proposition}

\begin{proof}
Let $\tau$ be a minimum-height spanning tree for $G$, rooted at $v_{1}$. \ A
depth-first search on $\tau$ uses $2n-2$\ steps. \ Let $S_{1}$\ be the set of
vertices visited by depth-first search in steps $1$ to $\delta_{G}$, $S_{2}%
$\ be those visited in steps $\delta_{G}+1$\ to $2\delta_{G}$, and so on.
\ Then%
\[
S_{1}\cup\cdots\cup S_{2n/\delta_{G}}=V\text{.}%
\]
Furthermore, for each $S_{j}$ there is a classical algorithm $A_{j}$, using at
most $3\delta_{G}$ steps, that starts at $v_{1}$, ends at $v_{1}$, and outputs
`$1$' if and only if $x_{i}=1$\ for some $v_{i}\in S_{j}$. \ Then we simply
perform Grover search at $v_{1}$\ over all $A_{j}$; since each iteration takes
$O\left(  \delta_{G}\right)  $\ steps and there are $O\left(  \sqrt
{2n/\delta_{G}}\right)  $ iterations, the number of steps is $O\left(
\sqrt{n\delta_{G}}\right)  $.
\end{proof}

The bound of Proposition \ref{upperdg}\ is tight:

\begin{theorem}
\label{lowerdg}For all $\delta$,\ there exists a graph $G$ with diameter
$\delta_{G}=\delta$\ such that%
\[
Q\left(  \operatorname*{OR},G\right)  =\Omega\left(  \sqrt{n\delta}\right)  .
\]

\end{theorem}

\begin{proof}
Let $G$ be a `starfish' with central vertex $v_{1}$\ and $M=2\left(
n-1\right)  /\delta$ legs $L_{1},\ldots,L_{M}$, each of length $\delta/2$ (see
Figure \ref{starfish}).%
\begin{figure}
[ptb]
\begin{center}
\includegraphics[
height=111.375pt,
width=125.375pt
]%
{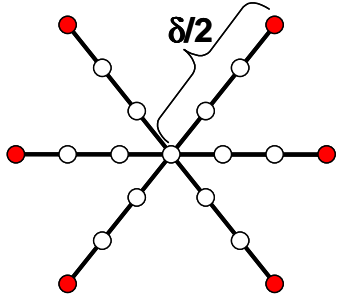}%
\caption{The `starfish' graph $G$. \ The marked item is at one of the tip
vertices.}%
\label{starfish}%
\end{center}
\end{figure}
We use the hybrid argument of Bennett et al. \cite{bbbv}. \ Suppose we run the
algorithm on the all-zero input $X_{0}$. \ Then define the \textit{query
magnitude} $\Gamma_{j}^{\left(  t\right)  }$\ to be the probability of finding
the robot in leg\ $L_{j}$ immediately after the $t^{th}$\ query:%
\[
\Gamma_{j}^{\left(  t\right)  }=\sum_{v_{i}\in L_{j}\,}\sum_{z\,}\left\vert
\alpha_{i,z}^{\left(  t\right)  }\left(  X_{0}\right)  \right\vert
^{2}\text{.}%
\]
Let $T$ be the total number of queries, and let $w=T/\left(  c\delta\right)  $
for some constant $0<c<1/2$. \ Clearly%
\[
\sum_{q=0}^{w-1}\sum_{j=1}^{M}\Gamma_{j}^{\left(  T-qc\delta\right)  }\leq
\sum_{q=0}^{w-1}1=w\text{.}%
\]
Hence there must exist a leg $L_{j^{\ast}}$\ such that
\[
\sum_{q=0}^{w-1}\Gamma_{j^{\ast}}^{\left(  T-qc\delta\right)  }\leq\frac{w}%
{M}=\frac{w\delta}{2\left(  n-1\right)  }.
\]
Let $v_{i^{\ast}}$ be the tip vertex of $L_{j^{\ast}}$, and let $Y$ be the
input which is $1$\ at $v_{i^{\ast}}$\ and $0$ elsewhere. \ Then let $X_{q}%
$\ be a hybrid input, which is $X_{0}$\ during queries $1$ to $T-qc\delta$,
but $Y$ during queries $T-qc\delta+1$ to $T$. \ Also, let%
\[
\left\vert \psi^{\left(  t\right)  }\left(  X_{q}\right)  \right\rangle
=\sum_{i,z}\alpha_{i,z}^{\left(  t\right)  }\left(  X_{q}\right)  \left\vert
v_{i},z\right\rangle
\]
be the algorithm's state after $t$ queries when run on $X_{q}$, and let%
\begin{align*}
D\left(  q,r\right)   &  =\left\Vert \left\vert \psi^{\left(  T\right)
}\left(  X_{q}\right)  \right\rangle -\left\vert \psi^{\left(  T\right)
}\left(  X_{r}\right)  \right\rangle \right\Vert _{2}^{2}\\
&  =\sum_{v_{i}\in G\,}\sum_{z\,}\left\vert \alpha_{i,z}^{\left(  T\right)
}\left(  X_{q}\right)  -\alpha_{i,z}^{\left(  T\right)  }\left(  X_{r}\right)
\right\vert ^{2}\text{.}%
\end{align*}
Then for all $q\geq1$,\ we claim that $D\left(  q-1,q\right)  \leq
4\Gamma_{j^{\ast}}^{\left(  T-qc\delta\right)  }$. \ For by unitarity, the
Euclidean distance between $\left\vert \psi^{\left(  t\right)  }\left(
X_{q-1}\right)  \right\rangle $\ and $\left\vert \psi^{\left(  t\right)
}\left(  X_{q}\right)  \right\rangle $\ can only increase as a result of
queries $T-qc\delta+1$\ through $T-\left(  q-1\right)  c\delta$. \ But no
amplitude from outside $L_{j^{\ast}}$\ can reach $v_{i^{\ast}}$\ during that
interval, since the distance is $\delta/2$\ and there are only $c\delta
<\delta/2$\ time steps. \ Therefore, switching from $X_{q-1}$\ to $X_{q}$\ can
only affect amplitude that is in $L_{j^{\ast}}$\ immediately after query
$T-qc\delta$:%
\begin{align*}
D\left(  q-1,q\right)   &  \leq\sum_{v_{i}\in L_{j^{\ast}}\,}\sum
_{z\,}\left\vert \alpha_{i,z}^{\left(  T-qc\delta\right)  }\left(
X_{q}\right)  -\left(  -\alpha_{i,z}^{\left(  T-qc\delta\right)  }\left(
X_{q}\right)  \right)  \right\vert ^{2}\\
&  =4\sum_{v_{i}\in L_{j^{\ast}}\,}\sum_{z\,}\left\vert \alpha_{i,z}^{\left(
T-qc\delta\right)  }\left(  X_{0}\right)  \right\vert ^{2}=4\Gamma_{j^{\ast}%
}^{\left(  T-qc\delta\right)  }.
\end{align*}
It follows that%
\[
\sqrt{D\left(  0,w\right)  }\leq\sum_{q=1}^{w}\sqrt{D\left(  q-1,q\right)
}\leq2\sum_{q=1}^{w}\sqrt{\Gamma_{j^{\ast}}^{\left(  T-qc\delta\right)  }}%
\leq2w\sqrt{\frac{\delta}{2\left(  n-1\right)  }}=\frac{T}{c}\sqrt{\frac
{2}{\delta\left(  n-1\right)  }}.
\]
Here the first inequality uses the triangle inequality, and the third uses the
Cauchy-Schwarz inequality. \ Now assuming the algorithm is correct we need
$D\left(  0,w\right)  =\Omega\left(  1\right)  $, which implies that
$T=\Omega\left(  \sqrt{n\delta}\right)  $.
\end{proof}

It is immediate that Theorem \ref{lowerdg}\ applies to $Z$-local unitaries as
well as $C$-local ones: that is, $Q^{Z}\left(  \operatorname*{OR},G\right)
=\Omega\left(  \sqrt{n\delta}\right)  $. \ We believe the theorem can be
extended to $H$-local unitaries as well, but a full discussion of this issue
would take us too far afield.

\section{Search on Grids\label{GRID}}

Let $\mathcal{L}_{d}\left(  n\right)  $\ be a $d$-dimensional grid graph of
size $n^{1/d}\times\cdots\times n^{1/d}$. \ That is, each vertex is specified
by $d$ coordinates\ $i_{1},\ldots,i_{d}\in\left\{  1,\ldots,n^{1/d}\right\}
$, and is connected to the at most $2d$ vertices obtainable by adding or
subtracting $1$ from a single coordinate (boundary vertices have fewer than
$2d$\ neighbors). \ We write simply $\mathcal{L}_{d}$\ when $n$ is clear from
context. \ In this section we present our main positive results: that
$Q\left(  \operatorname*{OR},\mathcal{L}_{d}\right)  =\Theta\left(  \sqrt
{n}\right)  $ for $d\geq3$, and $Q\left(  \operatorname*{OR},\mathcal{L}%
_{2}\right)  =O\left(  \sqrt{n}\operatorname*{polylog}n\right)  $\ for $d=2$.

Before proving these claims, let us develop some intuition by showing weaker
bounds, taking the case $d=2$ for illustration. \ Clearly $Q\left(
\operatorname*{OR},\mathcal{L}_{2}\right)  =O\left(  n^{3/4}\right)  $: we
simply partition $\mathcal{L}_{2}\left(  n\right)  $\ into $\sqrt{n}%
$\ subsquares, each a copy of $\mathcal{L}_{2}\left(  \sqrt{n}\right)  $. \ In
$5\sqrt{n}$\ steps, the robot can travel from the start vertex to any
subsquare $C$, search $C$\ classically for a marked vertex, and then return to
the start vertex. \ Thus, by searching all $\sqrt{n}$\ of the $C$'s in
superposition and applying Grover's algorithm, the robot can search the grid
in time $O\left(  n^{1/4}\right)  \times5\sqrt{n}=O\left(  n^{3/4}\right)  $.

Once we know that, we might as well partition $\mathcal{L}_{2}\left(
n\right)  $\ into $n^{1/3}$\ subsquares, each a copy of $\mathcal{L}%
_{2}\left(  n^{2/3}\right)  $. \ Searching any one of these subsquares by the
previous algorithm takes time $O\left(  \left(  n^{2/3}\right)  ^{3/4}\right)
=O\left(  \sqrt{n}\right)  $, an amount of time that also suffices to travel
to the subsquare and back from the start vertex. \ So using Grover's
algorithm, the robot can search $\mathcal{L}_{2}\left(  n\right)  $\ in time
$O\left(  \sqrt{n^{1/3}}\cdot\sqrt{n}\right)  =O\left(  n^{2/3}\right)  $.
\ We can continue recursively in this manner to make the running time approach
$O\left(  \sqrt{n}\right)  $. \ The trouble is that, with each additional
layer of recursion, the robot needs to repeat the search more often to
upper-bound the error probability. \ Using this approach, the best bounds we
could obtain are roughly $O\left(  \sqrt{n}\operatorname*{polylog}n\right)
$\ for $d\geq3$, or $\sqrt{n}2^{O\left(  \sqrt{\log n}\right)  }$\ for $d=2$.
\ In what follows, we use the amplitude amplification approach of Grover
\cite{grover:framework}\ and Brassard et al.\ \cite{bhmt}\ to improve these
bounds, in the case of a single marked vertex, to $O\left(  \sqrt{n}\right)
$\ for $d\geq3$ (Section \ref{D3}) and $O\left(  \sqrt{n}\log^{3/2}n\right)
$\ for $d=2$\ (Section \ref{D2}). \ Section \ref{MULTIPLE} generalizes these
results to the case of multiple marked vertices.

Intuitively, the reason the case $d=2$\ is special is that there, the diameter
of the grid is $\Theta\left(  \sqrt{n}\right)  $,\ which matches exactly the
time needed for Grover search. \ For $d\geq3$, by contrast, the robot can
travel across the grid in much less time than is needed to search it.

\subsection{Amplitude Amplification\label{AASEC}}

We start by describing amplitude amplification \cite{bhmt,grover:framework}, a
generalization of Grover search. \ Let $\mathcal{U}$ be a quantum algorithm
that, with probability $\epsilon$, outputs a correct answer together with a
witness that proves the answer correct. (For example, in the case of search,
the algorithm outputs a vertex label $i$ such that $x_{i}=1$.) \ Amplification
generates a new algorithm that calls $\mathcal{U}$ order $1/\sqrt{\epsilon}$
times, and that produces both a correct answer and a witness with probability
$\Omega\left(  1\right)  $. \ In particular, assume $\mathcal{U}$ starts in
basis state $\left\vert s\right\rangle $, and let $m$\ be a positive integer.
\ Then the amplification procedure works as follows:

\begin{enumerate}
\item[(1)] Set $\left\vert \psi_{0}\right\rangle =\mathcal{U}\left\vert
s\right\rangle $.

\item[(2)] For $i=1$ to $m$ set $\left\vert \psi_{i+1}\right\rangle
=\mathcal{U}S\mathcal{U}^{-1}W\left\vert \psi_{i}\right\rangle $, where

\begin{itemize}
\item $W$ flips the phase of basis state $\left\vert y\right\rangle $\ if and
only if $\left\vert y\right\rangle $\ contains a description of a correct
witness, and

\item $S$ flips the phase of basis state $\left\vert y\right\rangle $\ if and
only if $\left\vert y\right\rangle =\left\vert s\right\rangle $.
\end{itemize}
\end{enumerate}

We can decompose $\left\vert \psi_{0}\right\rangle $ as $\sin\alpha\left\vert
\Psi_{\operatorname*{succ}}\right\rangle +\cos\alpha\left\vert \Psi
_{\operatorname*{fail}}\right\rangle $, where $\left\vert \Psi
_{\operatorname*{succ}}\right\rangle $ is a superposition over basis states
containing a correct witness and $\left\vert \Psi_{\operatorname*{fail}%
}\right\rangle $ is a superposition over all other basis states. \ Brassard et
al. \cite{bhmt}\ showed the following:

\begin{lemma}
[\cite{bhmt}]\label{bhmtlem}$|\psi_{i}\rangle=\sin\left[  \left(  2i+1\right)
\alpha\right]  \left\vert \Psi_{\operatorname*{succ}}\right\rangle
+\cos\left[  \left(  2i+1\right)  \alpha\right]  \left\vert \Psi
_{\operatorname*{fail}}\right\rangle $.
\end{lemma}

If measuring $\left\vert \psi_{0}\right\rangle $ gives a correct witness with
probability $\epsilon$, then $\left\vert \sin\alpha\right\vert ^{2}=\epsilon$
and $\left\vert \alpha\right\vert \geq1/\sqrt{\epsilon}$. \ So taking
$m=O(1/\sqrt{\epsilon})$ yields $\sin\left[  \left(  2m+1\right)
\alpha\right]  \approx1$. \ For our algorithms, though, the multiplicative
constant under the big-O also matters. \ To upper-bound this constant, we
prove the following lemma.

\begin{lemma}
\label{Ampl}Suppose a quantum algorithm $\mathcal{U}$ outputs a correct answer
and witness with probability exactly $\epsilon$. \ Then by using $2m+1$ calls
to $\mathcal{U}$\ or $\mathcal{U}^{-1}$, where%
\[
m\leq\frac{\pi}{4\arcsin\sqrt{\epsilon}}-\frac{1}{2},
\]
we can output a correct answer and witness with probability at least%
\[
\left(  1-\frac{\left(  2m+1\right)  ^{2}}{3}\epsilon\right)  \left(
2m+1\right)  ^{2}\epsilon.
\]

\end{lemma}

\begin{proof}
We perform $m$ steps of amplitude amplification, which requires $2m+1$\ calls
to $\mathcal{U}$ or $\mathcal{U}^{-1}$. \ By Lemma \ref{bhmtlem}, this yields
the final state%
\[
\sin\left[  \left(  2m+1\right)  \alpha\right]  \left\vert \Psi
_{\operatorname*{succ}}\right\rangle +\cos\left[  \left(  2m+1\right)
\alpha\right]  \left\vert \Psi_{\operatorname*{fail}}\right\rangle .
\]
where $\alpha=\arcsin\sqrt{\epsilon}$. \ Therefore the success probability is
\begin{align*}
\sin^{2}\left[  \left(  2m+1\right)  \arcsin\sqrt{\epsilon}\right]   &
\geq\sin^{2}\left[  \left(  2m+1\right)  \sqrt{\epsilon}\right] \\
&  \geq\left(  \left(  2m+1\right)  \sqrt{\epsilon}-\frac{\left(  2m+1\right)
^{3}}{6}\epsilon^{3/2}\right)  ^{2}\\
&  \geq\left(  2m+1\right)  ^{2}\epsilon-\frac{\left(  2m+1\right)  ^{4}}%
{3}\epsilon^{2}.
\end{align*}
Here the first line uses the monotonicity of $\sin^{2}x$\ in the interval
$\left[  0,\pi/2\right]  $, and the second line uses the fact that $\sin x\geq
x-x^{3}/6$ for all $x\geq0$ by Taylor series expansion.
\end{proof}

Note that there is no need to uncompute any garbage left by $\mathcal{U}$,
beyond the uncomputation that happens \textquotedblleft
automatically\textquotedblright\ within the amplification procedure.

\subsection{Dimension At Least 3\label{D3}}

Our goal is the following:

\begin{theorem}
\label{sqrtsrchfull}If $d\geq3$, then $Q\left(  \operatorname*{OR}%
,\mathcal{L}_{d}\right)  =\Theta\left(  \sqrt{n}\right)  $.
\end{theorem}

In this section, we prove Theorem \ref{sqrtsrchfull}\ for the special case of
a unique marked vertex; then, in Sections \ref{MULTIPLE} and \ref{UNKNOWN}, we
will generalize to multiple marked vertices. \ Let $\operatorname*{OR}%
^{\left(  k\right)  }$\ be the problem of deciding whether there are no marked
vertices or exactly $k$ of them, given that one of these is true. \ Then:

\begin{theorem}
\label{sqrtsrch}If $d\geq3$, then $Q\left(  \operatorname*{OR}^{\left(
1\right)  },\mathcal{L}_{d}\right)  =\Theta\left(  \sqrt{n}\right)  $.
\end{theorem}

Choose constants $\beta\in\left(  2/3,1\right)  $\ and $\mu\in\left(
1/3,1/2\right)  $\ such that $\beta\mu>1/3$ (for example, $\beta=4/5$\ and
$\mu=5/11$ will work). \ Let $\ell_{0}$\ be a large positive integer; then for
all positive integers $R$, let $\ell_{R}=\ell_{R-1}\left\lceil \ell
_{R-1}^{1/\beta-1}\right\rceil $. \ Also let $n_{R}=\ell_{R}^{d}$. \ Assume
for simplicity that $n=n_{R}$\ for some $R$; in other words, that the
hypercube $\mathcal{L}_{d}\left(  n_{R}\right)  $\ to be searched has sides of
length $\ell_{R}$. \ Later we will remove this assumption.

Consider the following recursive algorithm $\mathcal{A}$. \ If $n=n_{0}$, then
search $\mathcal{L}_{d}\left(  n_{0}\right)  $ classically, returning $1$ if a
marked vertex is found and $0$ otherwise. \ Otherwise partition $\mathcal{L}%
_{d}\left(  n_{R}\right)  $\ into $n_{R}/n_{R-1}$\ subcubes, each one a copy
of $\mathcal{L}_{d}\left(  n_{R-1}\right)  $. \ Take the algorithm that
consists of picking a subcube $C$ uniformly at random, and then running
$\mathcal{A}$ recursively on $C$. \ Amplify this algorithm $\left(
n_{R}/n_{R-1}\right)  ^{\mu}$ times.

The intuition behind the exponents is that $n_{R-1}\approx n_{R}^{\beta}$, so
searching $\mathcal{L}_{d}\left(  n_{R-1}\right)  $\ should take about
$n_{R}^{\beta/2}$\ steps, which dominates the $n_{R}^{1/d}$ steps needed to
travel across the hypercube when $d\geq3$. \ Also, at level $R$ we want to
amplify a number of times that is less than $\left(  n_{R}/n_{R-1}\right)
^{1/2}$ by some polynomial amount, since full amplification would be
inefficient. \ The reason for the constraint $\beta\mu>1/3$\ will appear in
the analysis.

We now provide a more explicit description of $\mathcal{A}$, which shows that
it can be implemented using C-local unitaries and only a single bit of
workspace. \ At any time, the quantum robot's state will have the form
$\sum_{i,z}\alpha_{i,z}\left\vert v_{i},z\right\rangle $, where $v_{i}$\ is a
vertex of $\mathcal{L}_{d}\left(  n_{R}\right)  $ and $z$ is a single bit that
records whether or not a marked vertex has been found. \ Given a subcube $C$,
let $v\left(  C\right)  $ be the \textquotedblleft corner\textquotedblright%
\ vertex of $C$; that is, the vertex that is minimal in all $d$ coordinates.
\ Then the initial state when searching $C$\ will be $\left\vert v\left(
C\right)  ,0\right\rangle $. \ Beware, however, that \textquotedblleft initial
state\textquotedblright\ in this context\ just means the state $\left\vert
s\right\rangle $\ from Section \ref{AASEC}. \ Because of the way amplitude
amplification works, $\mathcal{A}$\ will often be invoked on $C$ with other
initial states, and even run in reverse.

For convenience, we will implement $\mathcal{A}$\ using a two-stage recursion:
given any subcube, the task of $\mathcal{A}$\ will be to amplify the result of
another procedure called $\mathcal{U}$, which in turn runs $\mathcal{A}%
$\ recursively on smaller subcubes. \ We will also use the conditional phase
flips $W$ and $S$ from Section \ref{AASEC}. \ For convenience, we write
$\mathcal{A}_{R},\mathcal{U}_{R},W_{R},S_{R}$\ to denote the level of
recursion that is currently active. \ Thus, $\mathcal{A}_{R}$\ calls
$\mathcal{U}_{R}$, which calls $\mathcal{A}_{R-1}$, which calls $\mathcal{U}%
_{R-1}$, and so on down to $\mathcal{A}_{0}$.

\begin{algorithm}
[$\mathcal{A}_{R}$]\label{alg12}\label{algw}Searches a subcube $C$ of size
$n_{R}$ for the marked vertex, and amplifies the result to have larger
probability. \ Default initial state: $\left\vert v\left(  C\right)
,0\right\rangle $.

\textbf{If }$R=0$\textbf{\ then:}

\begin{enumerate}
\item[(1)] Use classical C-local operations to visit all $n_{0}$\ vertices of
$C$ in any order. \ At each $v_{i}\in C$, use a query transformation to map
the state $\left\vert v_{i},z\right\rangle $\ to $\left\vert v_{i},z\oplus
x_{i}\right\rangle $.

\item[(2)] Return to $v\left(  C\right)  $.
\end{enumerate}

\textbf{If }$R\geq1$\textbf{\ then:}

\begin{enumerate}
\item[(1)] Let $m_{R}$ be the smallest integer such that $2m_{R}+1\geq\left(
n_{R}/n_{R-1}\right)  ^{\mu}$.

\item[(2)] Call $\mathcal{U}_{R}$.

\item[(3)] For $i=1$ to $m_{R}$, call $W_{R}$, then $\mathcal{U}_{R}^{-1}$,
then $S_{R}$, then $\mathcal{U}_{R}$. \ 
\end{enumerate}
\end{algorithm}

Suppose $\mathcal{A}_{R}$\ is run on the initial state $\left\vert v\left(
C\right)  ,0\right\rangle $, and let $C_{1},\ldots,C_{n_{R}/n_{0}}$\ be the
\textit{minimal subcubes} in $C$---meaning those of size $n_{0}$. \ Then the
final state after $\mathcal{A}_{R}$\ terminates should be%
\[
\frac{1}{\sqrt{n_{R}/n_{0}}}\sum_{i=1}^{n_{R}/n_{0}}\left\vert v\left(
C_{i}\right)  ,0\right\rangle
\]
if $C$ does not contain the marked vertex. \ Otherwise the final state should
have non-negligible overlap with $\left\vert v\left(  C_{i^{\ast}}\right)
,1\right\rangle $, where $C_{i^{\ast}}$\ is the minimal subcube in $C$ that
contains the marked vertex. \ In particular, if $R=0$, then the final state
should be $\left\vert v\left(  C\right)  ,1\right\rangle $\ if $C$ contains
the marked vertex, and $\left\vert v\left(  C\right)  ,0\right\rangle $\ otherwise.

The two phase-flip subroutines, $W_{R}$\ and $S_{R}$, are both trivial to
implement. \ To apply $W_{R}$, map each basis state $\left\vert v_{i}%
,z\right\rangle $\ to $\left(  -1\right)  ^{z}\left\vert v_{i},z\right\rangle
$. \ To apply $S_{R}$, map each $\left\vert v_{i},z\right\rangle $\ to
$-\left\vert v_{i},z\right\rangle $\ if $z=0$\ and $v_{i}=v\left(  C\right)
$\ for some subcube $C$ of size $n_{R}$,\ and to $\left\vert v_{i}%
,z\right\rangle $\ otherwise. \ Below we give pseudocode for $\mathcal{U}_{R}$.

\begin{algorithm}
[$\mathcal{U}_{R}$]\label{alg13}\label{alga}Searches a subcube $C$ of size
$n_{R}$\ for the marked vertex. \ Default initial state: $\left\vert v\left(
C\right)  ,0\right\rangle $.

\begin{enumerate}
\item[(1)] Partition $C$ into $n_{R}/n_{R-1}$\ smaller subcubes $C_{1}%
,\ldots,C_{n_{R}/n_{R-1}}$, each of size $n_{R-1}$.

\item[(2)] For all $j\in\left\{  1,\ldots,d\right\}  $, let $V_{j}$\ be the
set of corner vertices $v\left(  C_{i}\right)  $\ that differ from $v\left(
C\right)  $\ only in the first $j$ coordinates. \ Thus $V_{0}=\left\{
v\left(  C\right)  \right\}  $, and in general $\left\vert V_{j}\right\vert
=(\ell_{R}/\ell_{R-1})^{j}$. \ For $j=1$\ to $d$, let $\left\vert
V_{j}\right\rangle $ be the state%
\[
\left\vert V_{j}\right\rangle =\frac{1}{\ell_{R}^{j/2}}\sum_{v\left(
C_{i}\right)  \in V_{j}}\left\vert v\left(  C_{i}\right)  ,0\right\rangle
\]
Apply a sequence of transformations $Z_{1}$, $Z_{2}$, $\ldots$, $Z_{d}$ where
$Z_{j}$ is a unitary that maps $\left\vert V_{j-1}\right\rangle $\ to
$\left\vert V_{j}\right\rangle $\ by applying C-local\ unitaries that move
amplitude only along the $j^{th}$\ coordinate.

\item[(3)] Call $\mathcal{A}_{R-1}$\ recursively. \ (Note that this searches
$C_{1},\ldots,C_{n_{R}/n_{R-1}}$\ in superposition. \ Also, the required
amplification is performed for each of these subcubes automatically by step
(3) of $\mathcal{A}_{R-1}$.)
\end{enumerate}
\end{algorithm}

If $\mathcal{U}_{R}$\ is run on the initial state $\left\vert v\left(
C\right)  ,0\right\rangle $, then the final state should be%
\[
\frac{1}{\sqrt{n_{R}/n_{R-1}}}\sum_{i=1}^{n_{R}/n_{0}}\left\vert \phi
_{i}\right\rangle ,
\]
where $\left\vert \phi_{i}\right\rangle $\ is the correct final state when
$\mathcal{A}_{R-1}$\ is run on subcube $C_{i}$ with initial state $\left\vert
v\left(  C_{i}\right)  ,0\right\rangle $. \ A key point is that there is no
need for $\mathcal{U}_{R}$\ to call $\mathcal{A}_{R-1}$\ twice, once to
compute and once to uncompute---for the uncomputation is already built into
$\mathcal{A}_{R}$. \ \ This is what will enable us to prove an upper bound of
$O\left(  \sqrt{n}\right)  $\ instead of $O\left(  \sqrt{n}2^{R}\right)
=O\left(  \sqrt{n}\operatorname*{polylog}n\right)  $.

We now analyze the running time of $\mathcal{A}_{R}$.

\begin{lemma}
\label{d3lemma}$\mathcal{A}_{R}$ uses $O\left(  n_{R}^{\mu}\right)  $\ steps.
\end{lemma}

\begin{proof}
Let $T_{\mathcal{A}}\left(  R\right)  $\ and $T_{\mathcal{U}}\left(  R\right)
$\ be the total numbers of steps used by $\mathcal{A}_{R}$ and $\mathcal{U}%
_{R}$ respectively in searching $\mathcal{L}_{d}\left(  n_{R}\right)  $.
\ Then we have $T_{\mathcal{A}}\left(  0\right)  =O\left(  1\right)  $, and%
\begin{align*}
T_{\mathcal{A}}\left(  R\right)   &  \leq\left(  2m_{R}+1\right)
T_{\mathcal{U}}\left(  R\right)  +2m_{R}\\
T_{\mathcal{U}}\left(  R\right)   &  \leq dn_{R}^{1/d}+T_{\mathcal{A}}\left(
R-1\right)
\end{align*}
for all $R\geq1$. \ For $W_{R}$\ and $S_{R}$\ can both be implemented in a
single step, while $\mathcal{U}_{R}$\ uses $d\ell_{R}=dn_{R}^{1/d}$\ steps to
move the robot across the hypercube. \ Combining,%
\begin{align*}
T_{\mathcal{A}}\left(  R\right)   &  \leq\left(  2m_{R}+1\right)  \left(
dn_{R}^{1/d}+T_{\mathcal{A}}\left(  R-1\right)  \right)  +2m_{R}\\
&  \leq\left(  \left(  n_{R}/n_{R-1}\right)  ^{\mu}+2\right)  \left(
dn_{R}^{1/d}+T_{\mathcal{A}}\left(  R-1\right)  \right)  +\left(
n_{R}/n_{R-1}\right)  ^{\mu}+1\\
&  =O\left(  \left(  n_{R}/n_{R-1}\right)  ^{\mu}n_{R}^{1/d}\right)  +\left(
\left(  n_{R}/n_{R-1}\right)  ^{\mu}+2\right)  T_{\mathcal{A}}\left(
R-1\right) \\
&  =O\left(  \left(  n_{R}/n_{R-1}\right)  ^{\mu}n_{R}^{1/d}\right)  +\left(
n_{R}/n_{R-1}\right)  ^{\mu}T_{\mathcal{A}}\left(  R-1\right) \\
&  =O\left(  \left(  n_{R}/n_{R-1}\right)  ^{\mu}n_{R}^{1/d}+\left(
n_{R}/n_{R-2}\right)  ^{\mu}n_{R-1}^{1/d}+\cdots+\left(  n_{R}/n_{0}\right)
^{\mu}n_{1}^{1/d}\right) \\
&  =n_{R}^{\mu}\cdot O\left(  \frac{n_{R}^{1/d}}{n_{R-1}^{\mu}}+\frac
{n_{R-1}^{1/d}}{n_{R-2}^{\mu}}+\cdots+\frac{n_{1}^{1/d}}{n_{0}^{\mu}}\right)
\\
&  =n_{R}^{\mu}\cdot O\left(  n_{R}^{1/d-\beta\mu}+\cdots+n_{2}^{1/d-\beta\mu
}+n_{1}^{1/d-\beta\mu}\right) \\
&  =n_{R}^{\mu}\cdot O\left(  n_{R}^{1/d-\beta\mu}+\left(  n_{R}^{1/d-\beta
\mu}\right)  ^{1/\beta}+\cdots+\left(  n_{R}^{1/d-\beta\mu}\right)
^{1/\beta^{R-1}}\right) \\
&  =O\left(  n_{R}^{\mu}\right)  .
\end{align*}
Here the second line follows because $2m_{R}+1\leq\left(  n_{R}/n_{R-1}%
\right)  ^{\mu}+2$, the fourth because the $\left(  n_{R}/n_{R-1}\right)
^{\mu}$\ terms increase doubly exponentially, so adding $2$ to each will not
affect the asymptotics; the seventh because $n_{i}^{\mu}=\Omega\left(  \left(
n_{i+1}^{\mu}\right)  ^{\beta}\right)  $, the eighth because $n_{R-1}\leq
n_{R}^{\beta}$; and the last because $\beta\mu>1/3\geq1/d$, hence
$n_{1}^{1/d-\beta\mu}<1$.
\end{proof}

Next we need to lower-bound the success probability. \ Say that $\mathcal{A}%
_{R}$\ or $\mathcal{U}_{R}$\ \textquotedblleft succeeds\textquotedblright\ if
a measurement in the standard basis yields the result $\left\vert v\left(
C_{i^{\ast}}\right)  ,1\right\rangle $, where $C_{i^{\ast}}$\ is the minimal
subcube that contains the marked vertex. \ Of course, the marked vertex itself
can then be found in $n_{0}=O\left(  1\right)  $\ steps.

\begin{lemma}
\label{d3lemma2}Assuming there is a unique marked vertex, $\mathcal{A}_{R}$
succeeds with probability $\Omega\left(  1/n_{R}^{1-2\mu}\right)  $.
\end{lemma}

\begin{proof}
Let $P_{\mathcal{A}}\left(  R\right)  $\ and $P_{\mathcal{U}}\left(  R\right)
$\ be the success probabilities of $\mathcal{A}_{R}$ and $\mathcal{U}_{R}$
respectively when searching $\mathcal{L}_{d}\left(  n_{R}\right)  $. \ Then
clearly $P_{\mathcal{A}}\left(  0\right)  =1$, and $P_{\mathcal{U}}\left(
R\right)  =\left(  n_{R-1}/n_{R}\right)  P_{\mathcal{A}}\left(  R-1\right)
$\ for all $R\geq1$. \ So by Lemma \ref{Ampl},%
\begin{align*}
P_{\mathcal{A}}\left(  R\right)   &  \geq\left(  1-\frac{1}{3}\left(
2m_{R}+1\right)  ^{2}P_{\mathcal{U}}\left(  R\right)  \right)  \left(
2m_{R}+1\right)  ^{2}P_{\mathcal{U}}\left(  R\right) \\
&  =\left(  1-\frac{1}{3}\left(  2m_{R}+1\right)  ^{2}\frac{n_{R-1}}{n_{R}%
}P_{\mathcal{A}}\left(  R-1\right)  \right)  \left(  2m_{R}+1\right)
^{2}\frac{n_{R-1}}{n_{R}}P_{\mathcal{A}}\left(  R-1\right) \\
&  \geq\left(  1-\frac{1}{3}\left(  n_{R}/n_{R-1}\right)  ^{2\mu}\frac
{n_{R-1}}{n_{R}}P_{\mathcal{A}}\left(  R-1\right)  \right)  \left(
n_{R}/n_{R-1}\right)  ^{2\mu}\frac{n_{R-1}}{n_{R}}P_{\mathcal{A}}\left(
R-1\right) \\
&  \geq\left(  1-\frac{1}{3}\left(  n_{R-1}/n_{R}\right)  ^{1-2\mu}\right)
\left(  n_{R-1}/n_{R}\right)  ^{1-2\mu}P_{\mathcal{A}}\left(  R-1\right) \\
&  \geq\left(  n_{0}/n_{R}\right)  ^{1-2\mu}%
{\displaystyle\prod\limits_{r=1}^{R}}
\left(  1-\frac{1}{3}\left(  n_{R-1}/n_{R}\right)  ^{1-2\mu}\right) \\
&  \geq\left(  n_{0}/n_{R}\right)  ^{1-2\mu}%
{\displaystyle\prod\limits_{r=1}^{R}}
\left(  1-\frac{1}{3n_{R}^{\left(  1-\beta\right)  \left(  1-2\mu\right)  }%
}\right) \\
&  \geq\left(  n_{0}/n_{R}\right)  ^{1-2\mu}\left(  1-\sum_{r=1}^{R}\frac
{1}{3n_{R}^{\left(  1-\beta\right)  \left(  1-2\mu\right)  }}\right) \\
&  =\Omega\left(  1/n_{R}^{1-2\mu}\right)  .
\end{align*}
Here the third line follows because $2m_{R}+1\geq\left(  n_{R-1}/n_{R}\right)
^{\mu}$ and the function $x-\frac{1}{3}x^{2}$\ is nondecreasing in the
interval $\left[  0,1\right]  $;\ the fourth because $P_{\mathcal{A}}\left(
R-1\right)  \leq1$; the sixth because $n_{R-1}\leq n_{R}^{\beta}$; and the
last because $\beta<1$\ and $\mu<1/2$, the $n_{R}$'s increase doubly
exponentially, and $n_{0}$\ is sufficiently large.
\end{proof}

Finally, take $\mathcal{A}_{R}$ itself and amplify it to success probability
$\Omega\left(  1\right)  $ by running it $O(n_{R}^{1/2-\mu})$ times. \ This
yields an algorithm for searching $\mathcal{L}_{d}\left(  n_{R}\right)
$\ with overall running time $O\left(  n_{R}^{1/2}\right)  $, which implies
that $Q\left(  \operatorname*{OR}^{\left(  1\right)  },\mathcal{L}_{d}\left(
n_{R}\right)  \right)  =O\left(  n_{R}^{1/2}\right)  $.

All that remains is to handle values of $n$ that do not equal $n_{R}$\ for any
$R$. \ The solution is simple: first find the largest $R$ such that\ $n_{R}%
<n$. \ Then set $n^{\prime}=n_{R}\left\lceil n^{1/d}/\ell_{R}\right\rceil
^{d}$, and embed $\mathcal{L}_{d}\left(  n\right)  $\ into the larger
hypercube $\mathcal{L}_{d}\left(  n^{\prime}\right)  $. \ Clearly $Q\left(
\operatorname*{OR}^{\left(  1\right)  },\mathcal{L}_{d}\left(  n\right)
\right)  \leq Q\left(  \operatorname*{OR}^{\left(  1\right)  },\mathcal{L}%
_{d}\left(  n^{\prime}\right)  \right)  $. \ Also notice that $n^{\prime
}=O\left(  n\right)  $\ and that $n^{\prime}=O\left(  n_{R}^{1/\beta}\right)
=O\left(  n_{R}^{3/2}\right)  $. \ Next partition $\mathcal{L}_{d}\left(
n^{\prime}\right)  $\ into $n^{\prime}/n_{R}$\ subcubes, each a copy of
$\mathcal{L}_{d}\left(  n_{R}\right)  $. \ The algorithm will now have one
additional level of recursion, which chooses a subcube of $\mathcal{L}%
_{d}\left(  n^{\prime}\right)  $\ uniformly at random, runs $\mathcal{A}_{R}%
$\ on that subcube, and then amplifies the resulting procedure $\Theta\left(
\sqrt{n^{\prime}/n_{R}}\right)  $\ times. \ The total time is now%
\[
O\left(  \sqrt{\frac{n^{\prime}}{n_{R}}}\left(  \left(  n^{\prime}\right)
^{1/d}+n_{R}^{1/2}\right)  \right)  =O\left(  \sqrt{\frac{n^{\prime}}{n_{R}}%
}n_{R}^{1/2}\right)  =O\left(  \sqrt{n}\right)  ,
\]
while the success probability is $\Omega\left(  1\right)  $.\ \ This completes
Theorem \ref{sqrtsrch}.

\subsection{Dimension 2\label{D2}}

In the $d=2$\ case, the best we can achieve is the following:

\begin{theorem}
\label{sqrtsrch2full}$Q\left(  \operatorname*{OR},\mathcal{L}_{2}\right)
=O\left(  \sqrt{n}\log^{5/2}n\right)  $.
\end{theorem}

Again, we start with the single marked vertex case and postpone the general
case to Sections \ref{MULTIPLE} and \ref{UNKNOWN}.

\begin{theorem}
\label{sqrtsrch2}$Q\left(  \operatorname*{OR}^{\left(  1\right)  }%
,\mathcal{L}_{2}\right)  =O\left(  \sqrt{n}\log^{3/2}n\right)  $.
\end{theorem}

For $d\geq3$, we performed amplification on large (greater than $O\left(
1/n^{1-2\mu}\right)  $) probabilities only once, at the end. \ For $d=2$, on
the other hand, any algorithm that we construct with any nonzero success
probability will have running time $\Omega\left(  \sqrt{n}\right)  $, simply
because that is the diameter of the grid. \ If we want to keep the running
time $O\left(  \sqrt{n}\right)  $, then we can only perform $O\left(
1\right)  $ amplification steps at the end. \ Therefore we need to keep the
success probability relatively high throughout the recursion,\ meaning that we
suffer an increase in the running time, since amplification to high
probabilities is less efficient.

The procedures $\mathcal{A}_{R}$,\ $\mathcal{U}_{R}$, $W_{R}$, and $S_{R}$ are
identical to those in Section \ref{D3}; all that changes are the parameter
settings. \ For all integers $R\geq0$, we now let $n_{R}=\ell_{0}^{2R}$, for
some odd integer $\ell_{0}\geq3$\ to be set later. \ Thus, $\mathcal{A}_{R}%
$\ and $\mathcal{U}_{R}$\ search the square grid $\mathcal{L}_{2}\left(
n_{R}\right)  $ of size $\ell_{0}^{R}\times\ell_{0}^{R}$.\ \ Also, let
$m=\left(  \ell_{0}-1\right)  /2$; then $\mathcal{A}_{R}$\ applies $m$ steps
of amplitude amplification to $\mathcal{U}_{R}$.

We now prove the counterparts of Lemmas \ref{d3lemma}\ and \ref{d3lemma2} for
the two-dimensional case.

\begin{lemma}
\label{d2lemma}$\mathcal{A}_{R}$ uses $O\left(  R\ell_{0}^{R+1}\right)  $\ steps.
\end{lemma}

\begin{proof}
Let $T_{\mathcal{A}}\left(  R\right)  $\ and $T_{\mathcal{U}}\left(  R\right)
$\ be the time used by $\mathcal{A}_{R}$ and $\mathcal{U}_{R}$ respectively in
searching $\mathcal{L}_{2}\left(  n_{R}\right)  $. \ Then $T_{\mathcal{A}%
}\left(  0\right)  =1$, and for all $R\geq1$,%
\begin{align*}
T_{\mathcal{A}}\left(  R\right)   &  \leq\left(  2m+1\right)  T_{\mathcal{U}%
}\left(  R\right)  +2m,\\
T_{\mathcal{U}}\left(  R\right)   &  \leq2n_{R}^{1/2}+T_{\mathcal{A}}\left(
R-1\right)  .
\end{align*}
Combining,%
\begin{align*}
T_{\mathcal{A}}\left(  R\right)   &  \leq\left(  2m+1\right)  \left(
2n_{R}^{1/2}+T_{\mathcal{A}}\left(  R-1\right)  \right)  +2m\\
&  =\ell_{0}\left(  2\ell_{0}^{R}+T_{\mathcal{A}}\left(  R-1\right)  \right)
+\ell_{0}-1\\
&  =O\left(  \ell_{0}^{R+1}+\ell_{0}T_{\mathcal{A}}\left(  R-1\right)  \right)
\\
&  =O\left(  R\ell_{0}^{R+1}{}\right)  .
\end{align*}

\end{proof}

\begin{lemma}
\label{d2lemma2}$\mathcal{A}_{R}$ succeeds with probability $\Omega\left(
1/R\right)  $.
\end{lemma}

\begin{proof}
Let $P_{\mathcal{A}}\left(  R\right)  $\ and $P_{\mathcal{U}}\left(  R\right)
$\ be the success probabilities of $\mathcal{A}_{R}$ and $\mathcal{U}_{R}$
respectively when searching $\mathcal{L}_{2}\left(  n_{R}\right)  $. \ Then
$P_{\mathcal{U}}\left(  R\right)  =P_{\mathcal{A}}\left(  R-1\right)
/\ell_{0}^{2}$\ for all $R\geq1$. \ So by Lemma \ref{Ampl}, and using the fact
that $2m+1=\ell_{0}$,%
\begin{align*}
P_{\mathcal{A}}\left(  R\right)   &  \geq\left(  1-\frac{\left(  2m+1\right)
^{2}}{3}P_{\mathcal{U}}\left(  R\right)  \right)  \left(  2m+1\right)
^{2}P_{\mathcal{U}}\left(  R\right) \\
&  =\left(  1-\frac{\ell_{0}^{2}}{3}\frac{P_{\mathcal{A}}\left(  R-1\right)
}{\ell_{0}^{2}}\right)  \ell_{0}^{2}\frac{P_{\mathcal{A}}\left(  R-1\right)
}{\ell_{0}^{2}}\\
&  =P_{\mathcal{A}}\left(  R-1\right)  -\frac{1}{3}P_{\mathcal{A}}^{2}\left(
R-1\right) \\
&  =\Omega\left(  1/R\right)  .
\end{align*}
This is because $\Omega\left(  R\right)  $\ iterations of the map
$x_{R}:=x_{R-1}-\frac{1}{3}x_{R-1}^{2}$\ are needed to drop from (say)
$2/R$\ to $1/R$, and $x_{0}=P_{\mathcal{A}}\left(  0\right)  =1$ is greater
than $2/R$.
\end{proof}

We can amplify $\mathcal{A}_{R}$\ to success probability $\Omega\left(
1\right)  $\ by repeating it $O\left(  \sqrt{R}\right)  $\ times. \ This
yields an algorithm for searching $\mathcal{L}_{2}\left(  n_{R}\right)
$\ that uses $O\left(  R^{3/2}\ell_{0}^{R+1}\right)  =O\left(  \sqrt{n_{R}%
}R^{3/2}\ell_{0}\right)  $ steps in total. \ We can minimize this expression
subject to $\ell_{0}^{2R}=n_{R}$ by taking $\ell_{0}$\ to be constant and
$R$\ to be $\Theta\left(  \log n_{R}\right)  $, which yields $Q\left(
\operatorname*{OR}^{\left(  1\right)  },\mathcal{L}_{2}\left(  n_{R}\right)
\right)  =O\left(  \sqrt{n_{R}}\log n_{R}^{3/2}\right)  $. \ If $n$\ is not of
the form $\ell_{0}^{2R}$,\ then we simply find the smallest integer $R$ such
that $n<\ell_{0}^{2R}$, and embed $\mathcal{L}_{2}\left(  n\right)  $\ in the
larger grid $\mathcal{L}_{2}\left(  \ell_{0}^{2R}\right)  $. \ Since $\ell
_{0}$ is a constant, this increases the running time by at most a constant
factor. \ We have now proved Theorem \ref{sqrtsrch2}.

\subsection{Multiple Marked Items\label{MULTIPLE}}

What about the case in which there are multiple $i$'s\ with $x_{i}=1$? \ If
there are $k$ marked items (where $k$ need not be known in advance), then
Grover's algorithm can find a marked item with high probability in $O\left(
\sqrt{n/k}\right)  $\ queries, as shown by Boyer\ et al. \cite{bbht}. \ In our
setting, however, this is too much to hope for---since even if there are many
marked vertices, they might all be in a faraway part of the hypercube. \ Then
$\Omega\left(  n^{1/d}\right)  $ steps are needed, even if $\sqrt{n/k}%
<n^{1/d}$. \ Indeed, we can show a stronger lower bound. \ Recall that
$\operatorname*{OR}^{\left(  k\right)  }$\ is the problem of deciding whether
there are no marked vertices or exactly $k$ of them.

\begin{theorem}
\label{mmi}For all dimensions $d\geq2$,
\[
Q\left(  \operatorname*{OR}\nolimits^{\left(  k\right)  },\mathcal{L}%
_{d}\right)  =\Omega\left(  \frac{\sqrt{n}}{k^{1/2-1/d}}\right)  .
\]
Here, for simplicity, we ignore constant factors depending on $d$.
\end{theorem}

\begin{proof}
For simplicity, we assume that both $k^{1/d}$ and $\left(  n/3^{d}k\right)
^{1/d}$ are integers. \ (In the general case, we can just replace $k$ by
$\left\lceil k^{1/d}\right\rceil ^{d}$ and $n$ by the largest integer of the
form $\left(  3m\right)  ^{d}k$ which is less than $n$. \ This only changes
the lower bound by a constant factor depending on $d$.)

We use a hybrid argument almost identical to that of Theorem \ref{lowerdg}.
\ Divide $\mathcal{L}_{d}$ into $n/k$ subcubes, each having $k$ vertices and
side length $k^{1/d}$. \ Let $S$ be a regularly-spaced set of $M=n/\left(
3^{d}k\right)  $\ of these subcubes, so that any two subcubes in $S$ have
distance at least $2k^{1/d}$\ from one another. \ Then choose a subcube
$C_{j}\in S$\ uniformly at random and mark all $k$ vertices in $C_{j}$. \ This
enables us to consider each $C_{j}\in S$\ itself as a \textit{single} vertex
(out of $M$ in total), having distance at least $2k^{1/d}$\ to every other vertex.

More formally, given a subcube $C_{j}\in S$, let $\widetilde{C}_{j}$\ be the
set of vertices\ consisting of $C_{j}$\ and the $3^{d}-1$\ subcubes
surrounding it. \ (Thus, $\widetilde{C}_{j}$ is a subcube of side length
$3k^{1/d}$.) \ Then the query magnitude of $\widetilde{C}_{j}$\ after the
$t^{th}$\ query is%
\[
\Gamma_{j}^{\left(  t\right)  }=\sum_{v_{i}\in\widetilde{C}_{j}\,}\sum
_{z\,}\left\vert \alpha_{i,z}^{\left(  t\right)  }\left(  X_{0}\right)
\right\vert ^{2},
\]
where $X_{0}$\ is the all-zero input. \ Let $T$ be the number of queries, and
let $w=T/\left(  ck^{1/d}\right)  $ for some constant $c>0$. \ Then as in
Theorem \ref{lowerdg}, there must exist a subcube $\widetilde{C}_{j^{\ast}}%
$\ such that%
\[
\sum_{q=0}^{w-1}\Gamma_{j^{\ast}}^{\left(  T-qck^{1/d}\right)  }\leq\frac
{w}{M}=\frac{3^{d}kw}{n}.
\]
Let \thinspace$Y$ be the input which is $1$ in $C_{j^{\ast}}$\ and $0$
elsewhere; then let $X_{q}$\ be a hybrid input which is $X_{0}$\ during
queries $1$ to $T-qck^{1/d}$, but $Y$ during queries $T-qck^{1/d}+1$ to $T$.
\ Next let%
\[
D\left(  q,r\right)  =\sum_{v_{i}\in G\,}\sum_{z\,}\left\vert \alpha
_{i,z}^{\left(  T\right)  }\left(  X_{q}\right)  -\alpha_{i,z}^{\left(
T\right)  }\left(  X_{r}\right)  \right\vert ^{2}\text{.}%
\]
Then as in Theorem \ref{lowerdg}, for all $c<1$\ we have $D\left(
q-1,q\right)  \leq4\Gamma_{j^{\ast}}^{\left(  T-qck^{1/d}\right)  }$.\ For in
the $ck^{1/d}$\ queries from $T-qck^{1/d}+1$\ through $T-\left(  q-1\right)
ck^{1/d}$, no amplitude originating outside $\widetilde{C}_{j^{\ast}}$\ can
travel a distance $k^{1/d}$\ and thereby reach $C_{j^{\ast}}$. \ Therefore
switching from $X_{q-1}$\ to $X_{q}$\ can only affect amplitude that is in
$\widetilde{C}_{j^{\ast}}$\ immediately after query $T-qck^{1/d}$. \ It
follows that%
\[
\sqrt{D\left(  0,w\right)  }\leq\sum_{q=1}^{w}\sqrt{D\left(  q-1,q\right)
}\leq2\sum_{q=1}^{w}\sqrt{\Gamma_{j^{\ast}}^{\left(  T-qck^{1/d}\right)  }%
}\leq2w\sqrt{\frac{3^{d}k}{n}}=\frac{2\sqrt{3^{d}}k^{1/2-1/d}T}{c\sqrt{n}}.
\]
Hence $T=\Omega\left(  \sqrt{n}/k^{1/2-1/d}\right)  $\ for constant $d$, since
assuming the algorithm is correct we need $D\left(  0,w\right)  =\Omega\left(
1\right)  $.
\end{proof}

Notice that if $k\approx n$, then the bound of Theorem \ref{mmi} becomes
$\Omega\left(  n^{1/d}\right)  $ which is just the diameter of $\mathcal{L}%
_{d}$. \ Also, if $d=2$, then $1/2-1/d=0$ and the bound is simply
$\Omega\left(  \sqrt{n}\right)  $ independent of $k$. \ The bound of Theorem
\ref{mmi}\ can be achieved (up to a constant factor that depends on $d$) for
$d\geq3$, and nearly achieved for $d=2$. \ We first construct an algorithm for
the case when $k$ is known.

\begin{theorem}
\label{mmiub}\quad

\begin{enumerate}
\item[(i)] For $d\geq3$,%
\[
Q\left(  \operatorname*{OR}\nolimits^{\left(  k\right)  },\mathcal{L}%
_{d}\right)  =O\left(  \frac{\sqrt{n}}{k^{1/2-1/d}}\right)  .
\]

\item[(ii)] For $d=2$,%
\[
Q\left(  \operatorname*{OR}\nolimits^{\left(  k\right)  },\mathcal{L}%
_{2}\right)  =O\left(  \sqrt{n}\log^{3/2}n\right)  .
\]

\end{enumerate}
\end{theorem}

To prove Theorem \ref{mmiub}, we first divide $\mathcal{L}_{d}\left(
n\right)  $ into $n/\gamma$ subcubes, each of size $\gamma^{1/d}\times
\cdots\times\gamma^{1/d}$ (where $\gamma$ will be fixed later). \ Then in each
subcube, we choose one vertex uniformly at random.

\begin{lemma}
\label{rm}If $\gamma\geq k$, then the probability that exactly one marked
vertex is chosen is at least $k/\gamma-\left(  k/\gamma\right)  ^{2}$.
\end{lemma}

\begin{proof}
Let $x$ be a marked vertex. \ The probability that $x$ is chosen is $1/\gamma
$. \ Given that $x$ is chosen, the probability that one of the other marked
vertices, $y$, is chosen is $0$ if $x$ and $y$ belong to the same subcube, or
$1/\gamma$\ if they belong to different subcubes. \ Therefore, the probability
that $x$ alone is chosen is at least%
\[
\frac{1}{\gamma}\left(  1-\frac{k-1}{\gamma}\right)  \geq\frac{1}{\gamma
}\left(  1-\frac{k}{\gamma}\right)  .
\]
Since the events \textquotedblleft$x$ alone is chosen\textquotedblright\ are
mutually disjoint, we conclude that the probability that exactly one marked
vertex is chosen is at least $k/\gamma-\left(  k/\gamma\right)  ^{2}$.
\end{proof}

In particular, fix $\gamma$ so that $\gamma/3<k<2\gamma/3$; then Lemma
\ref{rm}\ implies that the probability of choosing exactly one marked vertex
is at least $2/9$. \ The algorithm is now as follows. \ As in the lemma,
subdivide $\mathcal{L}_{d}\left(  n\right)  $ into $n/\gamma$ subcubes and
choose one location at random from each. \ Then run the algorithm for the
unique-solution case (Theorem \ref{sqrtsrch} or \ref{sqrtsrch2}) on the chosen
locations only, as if they were vertices of $\mathcal{L}_{d}\left(
n/\gamma\right)  $.

The running time in the unique case was $O\left(  \sqrt{n/\gamma}\right)  $
for $d\geq3$ or
\[
O\left(  \sqrt{\frac{n}{\gamma}}\log^{3/2}\left(  n/\gamma\right)  \right)
=O\left(  \sqrt{\frac{n}{\gamma}}\log^{3/2}n\right)
\]
for $d=2$. \ However, each local unitary in the original algorithm now becomes
a unitary affecting two vertices $v$\ and $w$ in neighboring subcubes $C_{v}%
$\ and $C_{w}$. \ When placed side by side, $C_{v}$\ and $C_{w}$ form a
rectangular box of size $2\gamma^{1/d}\times\gamma^{1/d}\times\cdots
\times\gamma^{1/d}$. \ Therefore the distance between $v$ and $w$ is at most
$\left(  d+1\right)  \gamma^{1/d}$. \ It follows that each local unitary in
the original algorithm takes $O\left(  d\gamma^{1/d}\right)  $ time in the new
algorithm. \ For $d\geq3$, this results in an overall running time of%
\[
O\left(  \sqrt{\frac{n}{\gamma}}d\gamma^{1/d}\right)  =O\left(  d\frac
{\sqrt{n}}{\gamma^{1/2-1/d}}\right)  =O\left(  \frac{\sqrt{n}}{k^{1/2-1/d}%
}\right)  .
\]
For $d=2$\ we obtain%
\[
O\left(  \sqrt{\frac{n}{\gamma}}\gamma^{1/2}\log^{3/2}n\right)  =O\left(
\sqrt{n}\log^{3/2}n\right)  .
\]

\subsection{Unknown Number of Marked Items\label{UNKNOWN}}

We now show how to deal with an unknown $k$. \ Let $\operatorname*{OR}%
\nolimits^{\left(  \geq k\right)  }$\ be the problem of deciding whether there
are no marked vertices or \textit{at least} $k$ of them, given that one of
these is true.

\begin{theorem}
\label{mmiub1}\quad

\begin{enumerate}
\item[(i)] For $d\geq3$,%
\[
Q\left(  \operatorname*{OR}\nolimits^{\left(  \geq k\right)  },\mathcal{L}%
_{d}\right)  =O\left(  \frac{\sqrt{n}}{k^{1/2-1/d}}\right)  .
\]

\item[(ii)] For $d=2$,%
\[
Q\left(  \operatorname*{OR}\nolimits^{\left(  \geq k\right)  },\mathcal{L}%
_{2}\right)  =O\left(  \sqrt{n}\log^{5/2}n\right)  .
\]

\end{enumerate}
\end{theorem}

\begin{proof}
We use the straightforward `doubling' approach of Boyer et al. \cite{bbht}:

\begin{enumerate}
\item[(1)] For $j=0$\ to $\log_{2}\left(  n/k\right)  $

\begin{itemize}
\item Run the algorithm of Theorem \ref{mmiub} with subcubes of size
$\gamma_{j}=2^{j}k$.

\item If a marked vertex is found, then output $1$ and halt.
\end{itemize}

\item[(2)] Query a random vertex $v$, and output $1$ if $v$ is a marked vertex
and $0$ otherwise.
\end{enumerate}

Let $k^{\ast}\geq k$ be the number of marked vertices. $\ $If $k^{\ast}\leq
n/3$, then there exists a $j\leq\log_{2}\left(  n/k\right)  $ such that
$\gamma_{j}/3\leq k^{\ast}\leq2\gamma_{j}/3$. \ So Lemma \ref{rm} implies that
the $j^{th}$\ iteration of step (1) finds a marked vertex with probability at
least $2/9$. \ On the other hand, if $k^{\ast}\geq n/3$, then step (2) finds a
marked vertex with probability at least $1/3$. \ For $d\geq3$, the time used
in step (1) is at most%
\[
\sum_{j=0}^{\log_{2}\left(  n/k\right)  }\frac{\sqrt{n}}{\gamma_{j}^{1/2-1/d}%
}=\frac{\sqrt{n}}{k^{1/2-1/d}}\left[  \sum_{j=0}^{\log_{2}\left(  n/k\right)
}\frac{1}{2^{j\left(  1/2-1/d\right)  }}\right]  =O\left(  \frac{\sqrt{n}%
}{k^{1/2-1/d}}\right)  ,
\]
the sum in brackets being a decreasing geometric series. \ For $d=2$, the time
is $O\left(  \sqrt{n}\log^{5/2}n\right)  $, since each iteration takes
$O\left(  \sqrt{n}\log^{3/2}n\right)  $ time and there are at most $\log n$
iterations. \ In neither case does step (2) affect the bound, since $k\leq
n$\ implies that $n^{1/d}\leq\sqrt{n}/k^{1/2-1/d}$.
\end{proof}

Taking $k=1$ gives algorithms for unconstrained $\operatorname*{OR}$ with
running times $O(\sqrt{n})$ for $d\geq3$ and $O(\sqrt{n}\log^{5/2}n)$ for
$d=2$, thereby establishing Theorems \ref{sqrtsrchfull} and
\ref{sqrtsrch2full}.

\section{Search on Irregular Graphs\label{IRREG}}

In Section \ref{PREV}, we claimed that our divide-and-conquer approach has the
advantage of being \textit{robust}: it works not only for highly symmetric
graphs such as hypercubes, but for any graphs having comparable expansion
properties. \ Let us now substantiate this claim.

Say a family of connected graphs $\left\{  G_{n}=\left(  V_{n},E_{n}\right)
\right\}  $ is $d$\textit{-dimensional} if there exists a $\kappa>0$ such that
for all $n,\ell$ and $v\in V_{n}$,%
\[
\left\vert B\left(  v,\ell\right)  \right\vert \geq\min\left(  \kappa\ell
^{d},n\right)  ,
\]
where $B\left(  v,\ell\right)  $\ is the set of vertices having distance at
most $\ell$ from $v$ in $G_{n}$. \ Intuitively, $G_{n}$ is $d$-dimensional
(for $d\geq2$ an integer) if its expansion properties are at least as good as
those of the hypercube $\mathcal{L}_{d}\left(  n\right)  $.\footnote{In
general, it makes sense to consider non-integer $d$ as well.} \ It is
immediate that the diameter of $G_{n}$ is at most $\left(  n/\kappa\right)
^{1/d}$. \ Note, though, that $G_{n}$ might not be an expander graph in the
usual sense, since we have not required that every sufficiently small
\textit{set} of vertices has many neighbors.

Our goal is to show the following.

\begin{theorem}
\label{irregthm}If $G$ is $d$-dimensional, then

\begin{enumerate}
\item[(i)] For a constant $d>2$,%
\[
Q\left(  \operatorname*{OR},G\right)  =O\left(  \sqrt{n}%
\operatorname*{polylog}n\right)  .
\]

\item[(ii)] For $d=2$,%
\[
Q\left(  \operatorname*{OR},G\right)  =\sqrt{n}2^{O\left(  \sqrt{\log
n}\right)  }.
\]

\end{enumerate}
\end{theorem}

In proving part (i),\ the intuition is simple: we want to decompose $G$
recursively into subgraphs (called \textit{clusters}), which will serve the
same role as subcubes did in the hypercube case. \ The procedure is as
follows. \ For some constant $n_{1}>1$, first choose $\left\lceil
n/n_{1}\right\rceil $ vertices uniformly at random to be designated as
$1$-\textit{pegs}. \ Then form $1$\textit{-clusters} by assigning each vertex
in $G$ to its closest $1$-peg, as in a Voronoi diagram. \ (Ties are broken
randomly.) \ Let $v\left(  C\right)  $ be the peg of cluster $C$.\ \ Next,
split up any $1$-cluster $C$ with more than $n_{1}$\ vertices into
$\left\lceil \left\vert C\right\vert /n_{1}\right\rceil $ arbitrarily-chosen
$1$-clusters, each with size at most $n_{1}$ and with $v\left(  C\right)
$\ as its $1$-peg. \ Observe that%
\[
\sum_{i=1}^{\left\lceil n/n_{1}\right\rceil }\left\lceil \frac{\left\vert
C_{i}\right\vert }{n_{1}}\right\rceil \leq2\left\lceil \frac{n}{n_{1}%
}\right\rceil ,
\]
where $n=\left\vert C_{1}\right\vert +\cdots+\left\vert C_{\left\lceil
n/n_{1}\right\rceil }\right\vert $. \ Therefore, the splitting-up step can at
most double the number of clusters.

In the next iteration, set $n_{2}=n_{1}^{1/\beta}$, for some constant
$\beta\in\left(  2/d,1\right)  $. \ Choose $2\left\lceil n/n_{2}\right\rceil
$\ vertices uniformly at random as $2$-pegs. \ Then form $2$-clusters by
assigning each $1$-cluster $C$\ to the $2$-peg\ that is closest to the $1$-peg
$v\left(  C\right)  $. \ Given a $2$-cluster $C^{\prime}$, let $\left\vert
C^{\prime}\right\vert $\ be the number of $1$-clusters in $C^{\prime}$. \ Then
as before, split up any $C^{\prime}$\ with $\left\vert C^{\prime}\right\vert
>n_{2}/n_{1}$ into $\left\lceil \left\vert C^{\prime}\right\vert /\left(
n_{2}/n_{1}\right)  \right\rceil $\ arbitrarily-chosen $2$-clusters, each with
size at most $n_{2}/n_{1}$\ and with $v\left(  C^{\prime}\right)  $\ as its
$2$-peg. \ Continue recursively in this manner, setting $n_{R}=n_{R-1}%
^{1/\beta}$\ and choosing $2^{R-1}\left\lceil n/n_{R}\right\rceil $\ vertices
as $R$-pegs for each $R$. \ Stop at the maximum $R$ such that $n_{R}\leq n$.
\ For technical convenience, set $n_{0}=1$, and consider each vertex\ $v$ to
be the $0$-peg of the $0$-cluster $\left\{  v\right\}  $.

For $R\geq1$, define the \textit{radius} of an $R$-cluster $C$ to be the
maximum, over all $\left(  R-1\right)  $-clusters\ $C^{\prime}$\ in $C$, of
the distance from $v\left(  C\right)  $\ to $v\left(  C^{\prime}\right)  $.
\ Also, call an $R$-cluster \textit{good} if it has radius at most $\ell_{R}$,
where $\ell_{R}=\left(  \frac{2}{\kappa}n_{R}\ln n\right)  ^{1/d}$.

\begin{lemma}
\label{radiuslem}With probability $1-o\left(  1\right)  $ over the choice of
clusters, all clusters are good.
\end{lemma}

\begin{proof}
Let $v$\ be the $\left(  R-1\right)  $-peg of an $\left(  R-1\right)
$-cluster. \ Then $\left\vert B\left(  v,\ell\right)  \right\vert \geq
\kappa\ell^{d}$, where $B\left(  v,\ell\right)  $\ is the ball of radius
$\ell$\ about $v$. \ So the probability that $v$\ has distance greater than
$\ell_{R}$\ to the nearest $R$-peg\ is at most%
\[
\left(  1-\frac{\kappa\ell_{R}^{d}}{n}\right)  ^{\left\lceil n/n_{R}%
\right\rceil }\leq\left(  1-\frac{2\ln n}{n/n_{R}}\right)  ^{n/n_{R}}<\frac
{1}{n^{2}}.
\]
Furthermore, the total number of pegs is easily seen to be $O\left(  n\right)
$. \ It follows by the union bound that \textit{every} $\left(  R-1\right)
$-peg\ for \textit{every} $R$ has distance at most $\ell_{R}$\ to the nearest
$R$-peg, with probability $1-O\left(  1/n\right)  =1-o\left(  1\right)  $ over
the choice of clusters.
\end{proof}

At the end we have a tree of clusters, which can be searched recursively just
as in the hypercube case. Lemma \ref{radiuslem} gives us a guarantee on the
time needed to move a level down (from a peg of an $R$-cluster to a peg of an
$R-1$-cluster contained in it) or a level up. Also, let $K^{\prime}\left(
C\right)  $\ be the number of $\left(  R-1\right)  $-clusters in $R$-cluster
$C$; then $K^{\prime}\left(  C\right)  \leq K\left(  R\right)  $\ where
$K\left(  R\right)  =2\left\lceil n_{R}/n_{R-1}\right\rceil $. \ If
$K^{\prime}\left(  C\right)  <K\left(  R\right)  $, then place $K\left(
R\right)  -K^{\prime}\left(  C\right)  $\ \textquotedblleft
dummy\textquotedblright\ $\left(  R-1\right)  $-clusters in $C$, each of which
has $\left(  R-1\right)  $-peg\ $v\left(  C\right)  $. Now, every $R$-cluster
contains an equal number of $R-1$ clusters.

Our algorithm is similar to Section \ref{D3} but the basis states now have the
form $\left\vert v,z,C\right\rangle $,\ where $v$ is a vertex, $z$ is an
answer bit, and $C$\ is the label of the cluster currently being searched.
\ (Unfortunately, because multiple $R$-clusters can have the same peg, a
single auxiliary qubit no longer suffices.) \ 

The algorithm $\mathcal{A}_{R}$\ from Section \ref{D3}\ now does the
following, when invoked on the initial state $\left\vert v\left(  C\right)
,0,C\right\rangle $, where $C$\ is an $R$-cluster. \ If $R=0$, then
$\mathcal{A}_{R}$ uses a query transformation to prepare the state $\left\vert
v\left(  C\right)  ,1,C\right\rangle $\ if $v\left(  C\right)  $\ is the
marked vertex and $\left\vert v\left(  C\right)  ,0,C\right\rangle
$\ otherwise. \ If $R\geq1$ and $C$ is not a dummy cluster, then
$\mathcal{A}_{R}$\ performs $m_{R}$\ steps of amplitude amplification on
$\mathcal{U}_{R}$, where $m_{R}$ is the largest integer such that
$2m_{R}+1\leq\sqrt{n_{R}/n_{R-1}}$.\footnote{In the hypercube case, we
performed fewer amplifications in order to lower the running time from
$\sqrt{n}\operatorname*{polylog}n$\ to $\sqrt{n}$. \ Here, though, the
splitting-up step produces a $\operatorname*{polylog}n$\ factor anyway.} \ If
$C$ is a dummy cluster, then $\mathcal{A}_{R}$ does nothing for an appropriate
number of steps, and then returns that no marked item was found.

We now describe the subroutine $\mathcal{U}_{R}$, for $R\geq1$. \ When invoked
with $\left\vert v\left(  C\right)  ,0,C\right\rangle $ as its initial state,
$\mathcal{U}_{R}$\ first prepares a uniform superposition%
\[
\left\vert \phi_{C}\right\rangle =\frac{1}{\sqrt{K\left(  R\right)  }}%
\sum_{i=1}^{K\left(  R\right)  }\left\vert v\left(  C_{i}\right)
,0,C_{i}\right\rangle .
\]
It does this by first constructing a spanning tree $T$ for $C$, rooted at
$v\left(  C\right)  $\ and having minimal depth, and then moving amplitude
along the edges of $T$ so as to prepare $\left\vert \phi_{C}\right\rangle $.
\ After $\left\vert \phi_{C}\right\rangle $\ has been prepared, $\mathcal{U}%
_{R}$ then calls $\mathcal{A}_{R-1}$\ recursively, to search $C_{1}%
,\ldots,C_{K\left(  R\right)  }$\ in superposition and amplify the results.
\ Note that, because of the cluster labels, there is no reason why amplitude
being routed through $C$ should not pass through some other cluster
$C^{\prime}$\ along the way---but there is also no advantage in our analysis
for allowing this.

We now analyze the running time and success probability of $\mathcal{A}_{R}$.

\begin{lemma}
\label{irreglem}$\mathcal{A}_{R}$ uses $O\left(  \sqrt{n_{R}}\log
^{1/d}n\right)  $ steps, assuming that all clusters are good.
\end{lemma}

\begin{proof}
Let $T_{\mathcal{A}}\left(  R\right)  $\ and $T_{\mathcal{U}}\left(  R\right)
$\ be the time used by $\mathcal{A}_{R}$ and $\mathcal{U}_{R}$ respectively in
searching an $R$-cluster. \ Then we have%
\begin{align*}
T_{\mathcal{A}}\left(  R\right)   &  \leq\sqrt{n_{R}/n_{R-1}}T_{\mathcal{U}%
}\left(  R\right)  ,\\
T_{\mathcal{U}}\left(  R\right)   &  \leq\ell_{R}+T_{\mathcal{A}}\left(
R-1\right)
\end{align*}
with the base case $T_{\mathcal{A}}\left(  0\right)  =1$. \ Combining,%
\begin{align*}
T_{\mathcal{A}}\left(  R\right)   &  \leq\sqrt{n_{R}/n_{R-1}}\left(  \ell
_{R}+T_{\mathcal{A}}\left(  R-1\right)  \right) \\
&  \leq\sqrt{n_{R}/n_{R-1}}\ell_{R}+\sqrt{n_{R}/n_{R-2}}\ell_{R-1}%
+\cdots+\sqrt{n_{R}/n_{0}}\ell_{1}\\
&  =\sqrt{n_{R}}\cdot O\left(  \frac{\left(  n_{R}\ln n\right)  ^{1/d}}%
{\sqrt{n_{R-1}}}+\cdots+\frac{\left(  n_{1}\ln n\right)  ^{1/d}}{\sqrt{n_{0}}%
}\right) \\
&  =\sqrt{n_{R}}\left(  \ln^{1/d}n\right)  \cdot O\left(  n_{R}^{1/d-\beta
/2}+\cdots+n_{1}^{1/d-\beta/2}\right) \\
&  =\sqrt{n_{R}}\left(  \ln^{1/d}n\right)  \cdot O\left(  n_{1}^{1/d-\beta
/2}+\left(  n_{1}^{1/d-\beta/2}\right)  ^{1/\beta}+\cdots+\left(
n_{1}^{1/d-\beta/2}\right)  ^{\left(  1/\beta\right)  ^{R-1}}\right) \\
&  =O\left(  \sqrt{n_{R}}\log^{1/d}n\right)  ,
\end{align*}
where the last line holds because $\beta>2/d$ and therefore $n_{1}%
^{1/d-\beta/2}<1$.
\end{proof}

\begin{lemma}
\label{irreglem2}$\mathcal{A}_{R}$ succeeds with probability $\Omega\left(
1/\operatorname*{polylog}n_{R}\right)  $ in searching a graph of size
$n=n_{R}$, assuming there is a unique marked vertex.
\end{lemma}

\begin{proof}
For all $R\geq0$, let $C_{R}$\ be the $R$-cluster that contains the marked
vertex, and\ let $P_{\mathcal{A}}\left(  R\right)  $\ and $P_{\mathcal{U}%
}\left(  R\right)  $\ be the success probabilities of $\mathcal{A}_{R}$ and
$\mathcal{U}_{R}$ respectively when searching $C_{R}$. \ Then for all $R\geq
1$, we have $P_{\mathcal{U}}\left(  R\right)  =P_{\mathcal{A}}\left(
R-1\right)  /K\left(  R\right)  $, and therefore%
\begin{align*}
P_{\mathcal{A}}\left(  R\right)   &  \geq\left(  1-\frac{\left(
2m_{R}+1\right)  ^{2}}{3}P_{\mathcal{U}}\left(  R\right)  \right)  \left(
2m_{R}+1\right)  ^{2}P_{\mathcal{U}}\left(  R\right) \\
&  =\left(  1-\frac{\left(  2m_{R}+1\right)  ^{2}}{3}\cdot\frac{P_{\mathcal{A}%
}\left(  R-1\right)  }{K\left(  R\right)  }\right)  \left(  2m_{R}+1\right)
^{2}\frac{P_{\mathcal{A}}\left(  R-1\right)  }{K\left(  R\right)  }\\
&  =\Omega\left(  P_{\mathcal{A}}\left(  R-1\right)  \right) \\
&  =\Omega\left(  1/\operatorname*{polylog}n_{R}\right)  .
\end{align*}
Here the third line holds because $\left(  2m_{R}+1\right)  ^{2}\approx
n_{R}/n_{R-1}\approx K\left(  R\right)  /2$, and the last line because
$R=\Theta\left(  \log\log n_{R}\right)  $.
\end{proof}

Finally, we repeat $\mathcal{A}_{R}$\ itself $O(\operatorname*{polylog}n_{R})$
times, to achieve success probability $\Omega\left(  1\right)  $\ using
$O\left(  \sqrt{n_{R}}\operatorname*{polylog}n_{R}\right)  $\ steps in total.
\ Again, if $n$\ is not equal to $n_{R}$\ for any $R$, then we simply find the
largest $R$ such that $n_{R}<n$, and then add one more level of recursion that
searches a random $R$-cluster and amplifies the result $\Theta\left(
\sqrt{n/n_{R}}\right)  $\ times. \ The resulting algorithm uses $O\left(
\sqrt{n}\operatorname*{polylog}n\right)  $\ steps, thereby establishing part
(i) of Theorem \ref{irregthm}\ for the case of a unique marked vertex. \ The
generalization to multiple marked vertices is straightforward.

\begin{corollary}
\label{irregcor}If $G$ is $d$-dimensional for a constant $d>2$, then%
\[
Q\left(  \operatorname*{OR}\nolimits^{\left(  \geq k\right)  },G\right)
=O\left(  \frac{\sqrt{n}\operatorname*{polylog}\frac{n}{k}}{k^{1/2-1/d}%
}\right)  .
\]

\end{corollary}

\begin{proof}
Assume without loss of generality that $k=o\left(  n\right)  $, since
otherwise a marked item is trivially found in $O\left(  n^{1/d}\right)
$\ steps. \ As in Theorem \ref{mmiub1}, we give an algorithm $\mathcal{B}%
$\ consisting of $\log_{2}\left(  n/k\right)  +1$\ iterations. \ In iteration
$j=0$, choose $\left\lceil n/k\right\rceil $\ vertices $w_{1},\ldots
,w_{\left\lceil n/k\right\rceil }$\ uniformly at random. \ Then run the
algorithm for the unique marked vertex case, but instead of taking all
vertices in $G$ as $0$-pegs, take only $w_{1},\ldots,w_{\left\lceil
n/k\right\rceil }$. \ On the other hand, still choose the $1$-pegs, $2$-pegs,
and so on uniformly at random from among all vertices in $G$. \ For all $R$,
the number of $R$-pegs\ should be\ $\left\lceil \left(  n/k\right)
/n_{R}\right\rceil $. \ In general, in iteration $j$\ of $\mathcal{B}$,
choose\ $\left\lceil n/\left(  2^{j}k\right)  \right\rceil $\ vertices
$w_{1},\ldots,w_{\left\lceil n/\left(  2^{j}k\right)  \right\rceil }%
$\ uniformly at random, and then run the algorithm for a\ unique marked vertex
as if $w_{1},\ldots,w_{\left\lceil n/\left(  2^{j}k\right)  \right\rceil }%
$\ were the only vertices in the graph.

It is easy to see that, assuming there are $k$ or more marked vertices,
with\ probability $\Omega\left(  1\right)  $ there exists an iteration $j$
such that exactly one of $w_{1},\ldots,w_{\left\lceil n/\left(  2^{j}k\right)
\right\rceil }$\ is marked. \ Hence $\mathcal{B}$ succeeds with probability
$\Omega\left(  1\right)  $. \ It remains only to upper-bound $\mathcal{B}$'s
running time.

In\ iteration $j$, notice that Lemma \ref{radiuslem}\ goes through if we use
$\ell_{R}^{\left(  j\right)  }:=\left(  \frac{2}{\kappa}2^{j}kn_{R}\ln\frac
{n}{k}\right)  ^{1/d}$ instead of $\ell_{R}$. \ That is, with probability
$1-O\left(  k/n\right)  =1-o\left(  1\right)  $\ over the choice of clusters,
every $R$-cluster\ has radius at most $\ell_{R}^{\left(  j\right)  }$. \ So
letting $T_{\mathcal{A}}\left(  R\right)  $\ be the running time of
$\mathcal{A}_{R}$\ on an $R$-cluster, the recurrence in Lemma \ref{irreglem}
becomes%
\[
T_{\mathcal{A}}\left(  R\right)  \leq\sqrt{n_{R}/n_{R-1}}\left(  \ell
_{R}^{\left(  j\right)  }+T_{\mathcal{A}}\left(  R-1\right)  \right)
=O\left(  \sqrt{n_{R}}\left(  2^{j}k\log\left(  n/k\right)  \right)
^{1/d}\right)  ,
\]
which is%
\[
O\left(  \frac{\sqrt{n}\log^{1/d}\frac{n}{k}}{\left(  2^{j}k\right)
^{1/2-1/d}}\right)
\]
if $n_{R}=\Theta\left(  n/\left(  2^{j}k\right)  \right)  $. \ As usual, the
case where there is no $R$ such that $n_{R}=\Theta\left(  n/\left(
2^{j}k\right)  \right)  $\ is trivially handled by adding one more level of
recursion. \ If we factor in the $O\left(  1/\operatorname*{polylog}%
n_{R}\right)  $\ repetitions of $\mathcal{A}_{R}$\ needed to boost
the\ success probability to $\Omega\left(  1\right)  $, then the total running
time of iteration $j$\ is%
\[
O\left(  \frac{\sqrt{n}\operatorname*{polylog}\frac{n}{k}}{\left(
2^{j}k\right)  ^{1/2-1/d}}\right)  .
\]
Therefore $\mathcal{B}$'s running time is%
\[
O\left(  \sum_{j=0}^{\log_{2}\left(  n/k\right)  }\frac{\sqrt{n}%
\operatorname*{polylog}n}{\left(  2^{j}k\right)  ^{1/2-1/d}}\right)  =O\left(
\frac{\sqrt{n}\operatorname*{polylog}n}{k^{1/2-1/d}}\right)  .
\]

\end{proof}

For the $d=2$\ case, the best upper bound we can show is $\sqrt{n}2^{O\left(
\sqrt{\log n}\right)  }$. \ This is obtained by simply modifying
$\mathcal{A}_{R}$ to have a deeper recursion tree. \ Instead of taking
$n_{R}=n_{R-1}^{1/\mu}$ for some $\mu$, we take $n_{R}=2^{\sqrt{\log n}%
}n_{R-1}=2^{R\sqrt{\log n}}$, so that the total number of levels is
$\left\lceil \sqrt{\log n}\right\rceil $. \ Lemma \ref{radiuslem}\ goes
through without modification, while\ the recurrence for the running time
becomes%
\begin{align*}
T_{\mathcal{A}}\left(  R\right)   &  \leq\sqrt{n_{R}/n_{R-1}}\left(  \ell
_{R}+T_{\mathcal{A}}\left(  R-1\right)  \right) \\
&  \leq\sqrt{n_{R}/n_{R-1}}\ell_{R}+\sqrt{n_{R}/n_{R-2}}\ell_{R-1}%
+\cdots+\sqrt{n_{R}/n_{0}}\ell_{1}\\
&  =O\left(  2^{\sqrt{\log n}\left(  R/2\right)  }\sqrt{\ln n}+\cdots
+2^{\sqrt{\log n}\left(  R/2\right)  }\sqrt{\ln n}\right) \\
&  =\sqrt{n}2^{O\left(  \sqrt{\log n}\right)  }.
\end{align*}
Also, since the success probability decreases by at most a constant factor at
each level, we have that $P_{\mathcal{A}}\left(  R\right)  =2^{-O\left(
\sqrt{\log n}\right)  }$, and hence $2^{O\left(  \sqrt{\log n}\right)  }%
$\ amplification steps suffice to boost the success probability to
$\Omega\left(  1\right)  $. \ Handling multiple marked items adds an
additional factor of $\log n$, which is absorbed into $2^{O\left(  \sqrt{\log
n}\right)  }$. \ This completes Theorem \ref{irregthm}.

\subsection{Bits Scattered on a Graph\label{SCATTERED}}

In Section \ref{PHYS}, we discussed several ways to pack a given amount of
entropy into a spatial region of given dimensions. \ However, we said nothing
about how the entropy is \textit{distributed} within the region. \ It might be
uniform, or concentrated on the boundary, or distributed in some other way.
\ So we need to answer the following: suppose that in some graph, $h$\ out of
the $n$ vertices \textit{might} be marked, and we know which $h$ those are.
\ Then how much time is needed to determine whether any of the $h$ \textit{is}
marked? \ If the graph is the hypercube $\mathcal{L}_{d}$\ for $d\geq2$\ or is
$d$-dimensional for $d>2$, then the results of the previous sections imply
that $O\left(  \sqrt{n}\operatorname*{polylog}n\right)  $\ steps suffice.
\ However, we wish to use fewer steps, taking advantage of the fact that $h$
might be much smaller than $n$. \ Formally, suppose we are given a graph $G$
with $n$ vertices, of which \thinspace$h$ are potentially marked. \ Let
$\operatorname*{OR}\nolimits^{\left(  h,\geq k\right)  }$ be the problem of
deciding whether $G$ has no marked vertices or at least $k$ of them, given
that one of these is the case.

\begin{proposition}
\label{scatterlb}For all integer constants $d\geq2$, there exists a
$d$-dimensional graph $G$ such that%
\[
Q\left(  \operatorname*{OR}\nolimits^{\left(  h,\geq k\right)  },G\right)
=\Omega\left(  n^{1/d}\left(  \frac{h}{k}\right)  ^{1/2-1/d}\right)  .
\]

\end{proposition}

\begin{proof}
[Proof]Let $G$ be the $d$-dimensional hypercube $\mathcal{L}_{d}\left(
n\right)  $. \ Create $h/k$ subcubes of potentially marked vertices, each
having $k$ vertices and side length $k^{1/d}$. \ Space these subcubes out in
$\mathcal{L}_{d}\left(  n\right)  $\ so that the distance between any pair of
them is $\Omega\left(  \left(  nk/h\right)  ^{1/d}\right)  $. \ Then choose a
subcube $C$\ uniformly at random and mark all $k$ vertices in $C$. \ This
enables us to consider each subcube as a single vertex, having distance
$\Omega\left(  \left(  nk/h\right)  ^{1/d}\right)  $ to every other vertex.
\ The lower bound now follows by a hybrid argument essentially identical to
that of Theorem \ref{mmi}.
\end{proof}

In particular, if $d=2$\ then $\Omega\left(  \sqrt{n}\right)  $ time is always
needed, since the potentially marked vertices might all be far from the start
vertex. \ The lower bound of Proposition \ref{scatterlb}\ can be achieved up
to a polylogarithmic factor.

\begin{proposition}
\label{scatterthm}If $G$ is $d$-dimensional\ for a constant $d>2$, then%
\[
Q\left(  \operatorname*{OR}\nolimits^{\left(  h,\geq k\right)  },G\right)
=O\left(  n^{1/d}\left(  \frac{h}{k}\right)  ^{1/2-1/d}\operatorname*{polylog}%
\frac{h}{k}\right)  .
\]

\end{proposition}

\begin{proof}
Assume without loss of generality that $k=o\left(  h\right)  $, since
otherwise a marked item is trivially found.\ \ Use algorithm $\mathcal{B}%
$\ from Corollary \ref{irregcor}, with the following simple change. \ In
iteration $j$, choose\ $\left\lceil h/\left(  2^{j}k\right)  \right\rceil
$\ potentially marked vertices $w_{1},\ldots,w_{\left\lceil h/\left(
2^{j}k\right)  \right\rceil }$\ uniformly at random, and then run the
algorithm for a\ unique marked vertex as if $w_{1},\ldots,w_{\left\lceil
h/\left(  2^{j}k\right)  \right\rceil }$\ were the only vertices in the graph.
\ That is, take $w_{1},\ldots,w_{\left\lceil h/\left(  2^{j}k\right)
\right\rceil }$\ as $0$-pegs; then for all $R\geq1$, choose $\left\lceil
h/\left(  2^{j}kn_{R}\right)  \right\rceil $\ vertices of $G$ uniformly at
random as $R$-pegs. \ Lemma \ref{radiuslem}\ goes through if we use
$\widehat{\ell}_{R}^{\left(  j\right)  }:=\left(  \frac{2}{\kappa}\frac{n}%
{h}2^{j}kn_{R}\ln\frac{h}{k}\right)  ^{1/d}$ instead of $\ell_{R}$.\ \ So
following Corollary \ref{irregcor}, the running time of iteration $j$\ is now%
\[
O\left(  \sqrt{n_{R}}\left(  \frac{n}{h}2^{j}k\right)  ^{1/d}%
\operatorname*{polylog}\frac{h}{k}\right)  =O\left(  n^{1/d}\left(  \frac
{h}{2^{j}k}\right)  ^{1/2-1/d}\operatorname*{polylog}\frac{h}{k}\right)
\]
if $n_{R}=\Theta\left(  h/\left(  2^{j}k\right)  \right)  $. \ Therefore the
total running time is%
\[
O\left(  \sum_{j=0}^{\log_{2}\left(  h/k\right)  }n^{1/d}\left(  \frac
{h}{2^{j}k}\right)  ^{1/2-1/d}\operatorname*{polylog}\frac{h}{k}\right)
=O\left(  n^{1/d}\left(  \frac{h}{k}\right)  ^{1/2-1/d}\operatorname*{polylog}%
\frac{h}{k}\right)  .
\]

\end{proof}

Intuitively, Proposition \ref{scatterthm}\ says that the worst case for search
occurs when the $h$ potential marked vertices are scattered evenly throughout
the graph.

\section{Application to Disjointness\label{APPL}}

In this section we show how our results can be used to strengthen a seemingly
unrelated result in quantum computing. \ Suppose Alice has a string
$X=x_{1}\ldots x_{n}\in\left\{  0,1\right\}  ^{n}$, and Bob has a string
$Y=y_{1}\ldots y_{n}\in\left\{  0,1\right\}  ^{n}$. \ In the
\textit{disjointness problem}, Alice and Bob must decide with high probability
whether there exists an $i$ such that $x_{i}=y_{i}=1$, using as few bits of
communication as possible. \ Buhrman, Cleve, and Wigderson \cite{bcw}%
\ observed that in the quantum setting, Alice and Bob can solve this problem
using only $O\left(  \sqrt{n}\log n\right)  $\ qubits of communication. \ This
was subsequently improved by H\o yer and de Wolf \cite{hoyerdewolf}\ to
$O\left(  \sqrt{n}c^{\log^{\ast}n}\right)  $, where $c$\ is a constant and
$\log^{\ast}n$\ is the iterated logarithm function. \ Using the search
algorithm of Theorem \ref{sqrtsrchfull}, we can improve this to $O\left(
\sqrt{n}\right)  $, which matches the celebrated $\Omega\left(  \sqrt
{n}\right)  $\ lower bound of Razborov \cite{razborov:cc}.

\begin{theorem}
The quantum communication complexity of the disjointness problem is $O\left(
\sqrt{n}\right)  $.
\end{theorem}

\begin{proof}
The protocol is as follows. \ Alice and Bob both store their inputs in a $3$-D
cube $\mathcal{L}_{3}\left(  n\right)  $ (Figure \ref{alicebob}); that is,
they let $x_{jkl}=x_{i}$\ and $y_{jkl}=y_{i}$, where $i=n^{2/3}j+n^{1/3}%
k+l+1$\ and $j,k,l\in\left\{  0,\ldots,n^{1/3}-1\right\}  $.%
\begin{figure}
[ptb]
\begin{center}
\includegraphics[
height=1.0652in,
width=3.6073in
]%
{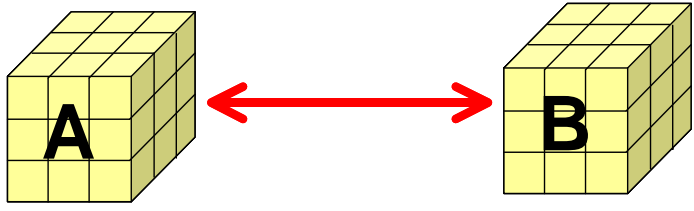}%
\caption{Alice and Bob `synchronize' locations on their respective cubes.}%
\label{alicebob}%
\end{center}
\end{figure}
To decide whether there exists a $\left(  j,k,l\right)  $\ with $x_{jkl}%
=y_{jkl}=1$, Alice simply runs our search algorithm for an unknown number of
marked items.  If the search algorithm is in the state
\[
\sum\alpha_{j,k,l,z}\left\vert v_{jkl},z\right\rangle ,
\]
then the joint state of Alice and Bob will be
\begin{equation}
\sum\alpha_{j,k,l,z,c}\left\vert v_{jkl}\right\rangle \otimes\left\vert
z\right\rangle \otimes\left\vert c\right\rangle \otimes\left\vert
v_{jkl}\right\rangle ,\label{theform}%
\end{equation}
where Alice holds the first $\left\vert v_{jkl}\right\rangle $ and $\left\vert
z\right\rangle $, Bob holds the second $\left\vert v_{jkl}\right\rangle $, and
$|c\rangle$ is the communication channel. \ Thus, whenever Alice is at
location $\left(  j,k,l\right)  $\ of her cube, Bob is at location $\left(
j,k,l\right)  $\ of his cube.

\begin{enumerate}
\item[(1)] To simulate a query, Alice sends $|z\rangle$ and an auxiliary qubit
holding $x_{jkl}$ to Bob. \ Bob performs $|z\rangle\rightarrow|z\oplus
y_{jkl}\rangle$, conditional on $x_{jkl}=1$. \ He then returns both bits to
Alice, and finally Alice returns the auxiliary qubit to the $\left\vert
0\right\rangle $ state by exclusive-OR'ing it with $x_{jkl}$.

\item[(2)] To simulate a non-query transformation that does not change
$\left\vert v_{jkl}\right\rangle $, Alice just performs it herself.

\item[(3)] By examining Algorithms \ref{alg12} and \ref{alg13}, we see that
there are two transformations that change $\left\vert v_{jkl}\right\rangle $.
\ We deal with them separately.

First, step 1 of Algorithm \ref{alg12} uses a classical $C$-local
transformation $\left\vert v_{j,k,l}\right\rangle \rightarrow|v_{j^{\prime
},k^{\prime},l^{\prime}}\rangle$. \ This transformation can be simulated by
Alice and Bob each separately applying $|v_{j,k,l}\rangle\rightarrow
|v_{j^{\prime},k^{\prime},l^{\prime}}\rangle$.

Second, step 2 of Algorithm \ref{alg13} applies transformations $Z_{1}$,
$Z_{2}$, and $Z_{3}$. \ For brevity, we restrict ourselves to discussing
$Z_{1}$. \ This transformation maps an initial state $\left\vert
v_{j,k,l},0\right\rangle $ to a uniform superposition over $|v_{j^{\prime
},k,l},0\rangle$ for all $\left(  j^{\prime},k,l\right)  $ lying in the same
$C_{i}$ as $\left(  j,k,l\right)  $. \ We can decompose this into a sequence
of transformations mapping $|v_{j^{\prime},k,l}\rangle$ to $\alpha
|v_{j^{\prime},k,l}\rangle+\beta|v_{j^{\prime}+1,k,l}\rangle$ for some
$\alpha$, $\beta$. \ This can be implemented in three steps, using an
auxiliary qubit. \ The auxiliary qubit is initialized to $\left\vert
0\right\rangle $ and is initially held by Alice. \ At the end, the auxiliary
qubit is returned to $\left\vert 0\right\rangle $. \ The sequence of
transformations is
\begin{align*}
|v_{j^{\prime},k,l}\rangle\left\vert 0\right\rangle |v_{j^{\prime},k,l}\rangle
& \rightarrow\alpha|v_{j^{\prime},k,l}\rangle\left\vert 0\right\rangle
|v_{j^{\prime},k,l}\rangle+\beta|v_{j^{\prime},k,l}\rangle\left\vert
1\right\rangle |v_{j^{\prime},k,l}\rangle\\
& \rightarrow\alpha|v_{j^{\prime},k,l}\rangle\left\vert 0\right\rangle
|v_{j^{\prime},k,l}\rangle+\beta|v_{j^{\prime},k,l}\rangle\left\vert
1\right\rangle |v_{j^{\prime}+1,k,l}\rangle\\
& \rightarrow\alpha|v_{j^{\prime},k,l}\rangle\left\vert 0\right\rangle
|v_{j^{\prime},k,l}\rangle\beta|v_{j^{\prime},k,l}\rangle\left\vert
0\right\rangle |v_{j^{\prime}+1,k,l}\rangle.
\end{align*}
The first transformation is performed by Alice who then sends the auxiliary
qubit to Bob. \ The second transformation is performed by Bob, who then sends
the auxiliary qubit back to Alice, who performs the third transformation.
\end{enumerate}

Since the algorithm uses $O\left(  \sqrt{n}\right)  $\ steps, and each step is
simulated using a constant amount of communication, the number of qubits
communicated in the disjointness protocol is therefore also $O\left(  \sqrt
{n}\right)  $.
\end{proof}

\section{Open Problems\label{OPEN}}

As discussed in Section \ref{LOCAL}, a salient open problem raised by this
work is to prove relationships among Z-local, C-local, and H-local unitary
matrices. \ In particular, can any Z-local or H-local unitary be approximated
by a product of a small number of C-local unitaries? \ Also, is it true that
$Q\left(  f,G\right)  =\Theta\left(  Q^{Z}\left(  f,G\right)  \right)
=\Theta\left(  Q^{H}\left(  f,G\right)  \right)  $\ for all $f,G$?

A second problem is to obtain interesting lower bounds in our model. \ For
example, let $G$\ be a $\sqrt{n}\times\sqrt{n}$\ grid, and suppose $f\left(
X\right)  =1$\ if and only if every row of $G$ contains a vertex $v_{i}$\ with
$x_{i}=1$. \ Clearly $Q\left(  f,G\right)  =O\left(  n^{3/4}\right)  $, and we
conjecture that this is optimal. \ However, we were unable to show any lower
bound better than $\Omega\left(  \sqrt{n}\right)  $.

Finally, what is the complexity of finding a unique marked vertex on a 2-D
square grid? \ As mentioned in Section \ref{PREV}, Ambainis, Kempe, and Rivosh
\cite{akr}\ showed that $Q\left(  \operatorname*{OR}^{\left(  1\right)
},\mathcal{L}_{2}\right)  =O\left(  \sqrt{n}\log n\right)  $. \ Can the
remaining factor of $\log n$\ be removed?

\section{Acknowledgments}

We thank Andrew Childs, Julia Kempe, Neil Shenvi, and Ronald de Wolf for
helpful conversations; Jakob Bekenstein, Raphael Bousso, and John Preskill for
discussions relevant to Section \ref{PHYS}; and the anonymous reviewers for
comments on the manuscript and a simpler proof of Lemma \ref{rm}.

\bibliographystyle{plain}
\bibliography{thesis}

\end{document}